# Universal statistical laws governing culinary design

Ganesh Bagler[1,2,3,6*], Gopal Krishna Tewari[4†], Aditya Raj Yadav[4†], Akshat Singh[5†], Pranay Bansal[5†], Ujjval Dargar[5†], Mansi Goel[1,2,3†], and Madhvi Kumari Sinha[1,2,3†]

[1]Department of Computational Biology
[2]Infosys Center for Artificial Intelligence
[3]Center of Excellence in Healthcare
[4]Department of Mathematics
[5]Department of Computer Science
Indraprastha Institute of Information Technology Delhi (IIIT-Delhi), New Delhi 110020 India.
[6]Foodoscope Technologies Private Limited, New Delhi 110048 India.

*Corresponding Author: bagler@iiitd.ac.in
†Equal contribution

**Cooking is a cultural expression of human creativity[1,2] that transcends geography and time through the orchestration of ingredients and techniques[3–6], much like languages do through words and syntax[7,8]. Yet, beneath the apparent diversity of culinary traditions, whether recipes obey statistical laws comparable to those of other symbolic systems remains unknown. Here we analyze a large corpus of traditional recipes spanning global cuisines, annotated using a state-of-the-art named entity recognition algorithm into ingredients, cooking techniques, utensils, and other culinary attributes[9–14]. We find that ingredient usage exhibits Zipf-like rank–frequency scaling, that culinary diversity grows sublinearly with corpus size in accordance with Heaps' law, and that recipe complexity follows Menzerath–Altmann-type relations between the number and average information of constituent units. Consistent with observations in packaged foods[15], macronutrient concentrations across recipes also display a log-normal signature. Minimal generative models based on preferential reuse, constrained sampling, and incremental modification recapitulate these regularities, suggesting generic processes that shape recipe architecture across cultures. Together, these findings establish recipes as a compositional symbolic system in which complex structure emerges from simple, constrained generative processes.**

Cooking is one of the most creative and quintessential human endeavors[1,2]. Culinary traditions have evolved through the imaginative composition of ingredients and the inventive transformation of raw materials into nourishing and pleasurable dishes. This creativity is expressed not only in the choice of ingredients, but also in the ways they are combined and processed. At its core, a recipe is a structured act of design, a sequence of decisions that balance taste, nutrition, culture, availability, and technique. One of the deepest questions one can ask about such creative systems is whether there exist inherent laws governing how these compositions emerge.

Recipes encode how ingredients, processing steps, and techniques are orchestrated into dishes that are palatable, nutritious, and culturally meaningful[1]. In this sense, the act of composing a recipe bears a striking resemblance to the formation of sentences in a natural language[7–9]: discrete units are combined according to implicit rules shaped by shared conventions and constraints. Despite the striking analogy, whether recipes, as compositional symbolic systems, exhibit statistical structure comparable to language remains largely unexplored.

We propose that recipes can be viewed as compositional symbolic systems, in which discrete culinary units such as ingredients, techniques are combined under implicit constraints to produce structured outcomes. In this view, recipes are not merely cultural artifacts but instances of a broader class of systems that, like language and other symbolic constructs, exhibit measurable statistical organization.

In linguistics and quantitative social science, a family of empirical scaling relations has been documented across diverse corpora[16]. Zipf's law describes the heavy-tailed distribution of word frequencies, with a small number of words used extremely often and a long tail of rare terms[17]. Heaps' law captures how vocabulary size grows sublinearly with corpus size, reflecting diminishing returns in the discovery of new words[18]. The Menzerath–Altmann law relates the number of units in a structure to the average size of those units, revealing systematic trade-offs in how complex entities are composed[19,20]. Similar heavy-tailed distributions and scaling laws have been observed in domains such as city sizes[21,22], scientific citations[23], and income distributions[24], and are often interpreted as signatures of underlying generative mechanisms including preferential attachment[25], constraint-driven evolution[18,26], and hierarchical organization[19,20].

Culinary systems share essential features with these domains. Recipes draw from a large but finite repertoire of ingredients, cooking techniques, preprocessing methods, and utensils[9,10]. New dishes rarely emerge de novo; instead, they are typically generated by modifying or recombining existing recipes, much as new sentences or cultural artefacts arise through variation on prior forms[27–29]. Ingredient and technique choices are constrained by physiology, perception, economics, and culture, narrowing the effectively used subset of possibilities. Over historical time, cuisines diversify and specialize, giving rise to rich regional variations while preserving recognizable structural motifs[3–6]. These characteristics suggest that recipes may be governed by statistical regularities reflecting both human cognitive constraints and broader socio-ecological factors.

Previous work at the intersection of data science and gastronomy has demonstrated that culinary data exhibit non-trivial organization[3–6,30–34] in ingredient usage, flavor pairing patterns, and cross-cultural differences in recipe design. Network analyses of ingredient co-occurrence have revealed flavor-complementarity principles in specific cuisines[3–6]. Computational studies have proposed algorithmic methods for generating novel dishes[35–39]. However, these efforts have largely focused on cuisine-specific phenomena (such as ingredient pairings) rather than asking whether recipes, as symbolic compositional symbolic systems, obey general statistical laws analogous to those found in language[16] and other complex systems[21–24].

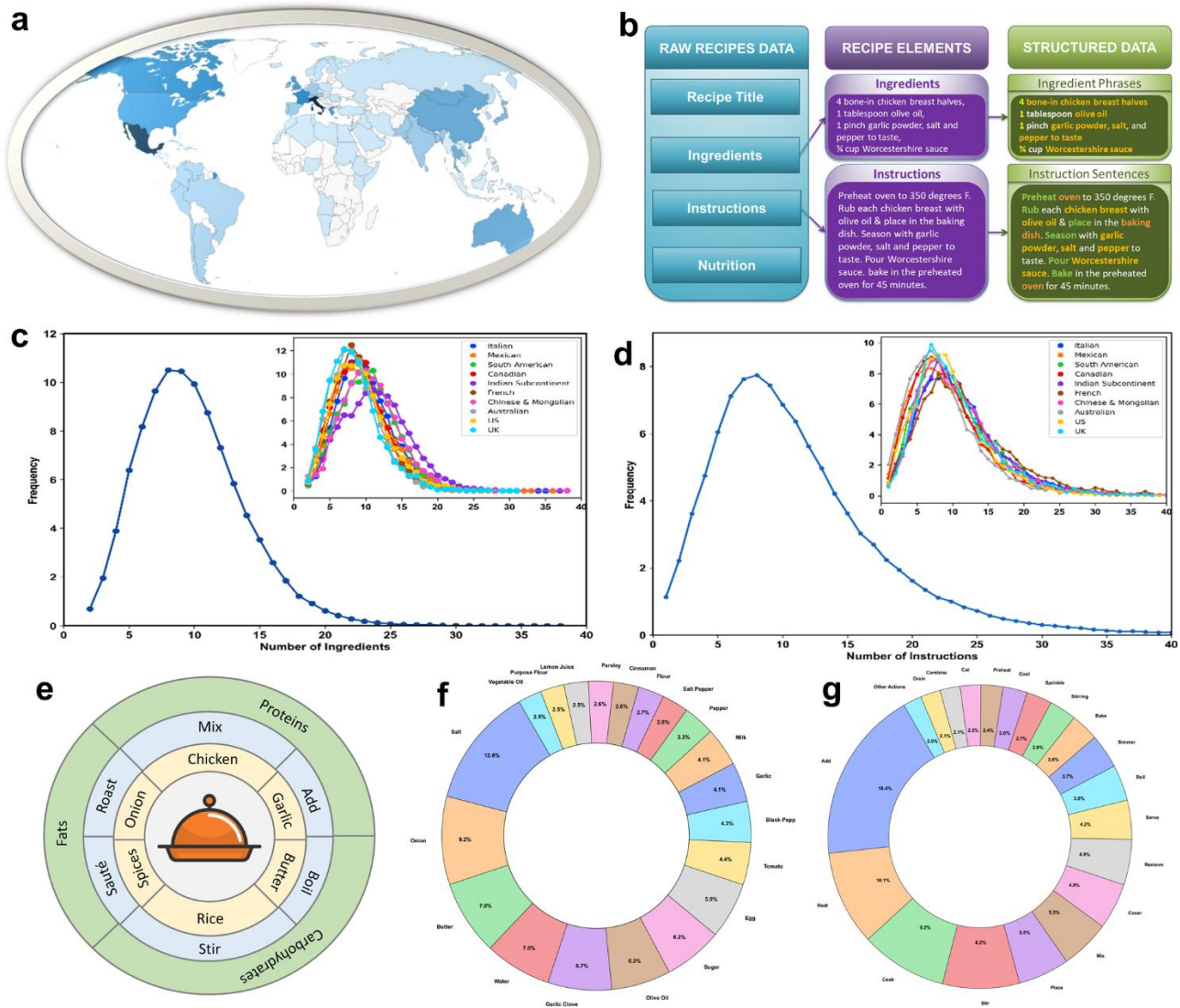

*Fig. 1 | Global culinary corpus and structured annotation pipeline.* Overview of the dataset, preprocessing framework, and key statistical properties of the global recipe corpus. The dataset comprises 118,083 recipes curated from 26 cuisines spanning diverse geographical, cultural, and climatic regions. **a,** Geographic distribution of recipes: The corpus encompasses major culinary traditions across the world. The spatial distribution highlights broad global coverage, ensuring representation of diverse ingredient repertoires and cooking practices. **b,** Structured annotation pipeline: Raw recipe text is processed using a named entity recognition framework to extract structured culinary entities. Ingredients phrases are decomposed into standardized attributes (name, quantity, unit, state, preprocessing), while instructions are parsed into discrete action phrases capturing cooking techniques and utensils. **c,** Distribution of recipe size: The number of ingredients per recipe exhibits a unimodal distribution, peaking around 8–12 ingredients, indicating a characteristic scale. Inset shows cuisine-specific distributions. **d,** Distribution of preparation complexity: The number of instruction steps per recipe also follows a unimodal distribution, with a median of approximately nine steps. Cuisine-specific distributions (inset) reveal similar structural organization of procedural complexity across diverse food cultures. **e,** Schematic representation of culinary composition: Recipes are

*conceptualized as structured compositions of ingredients and cooking actions. Ingredients are combined and transformed through operations such as mixing, boiling, and roasting, illustrating the interplay between composition and process in generating culinary outcomes. **f**, Ingredient usage distribution: The most frequently used ingredients across the global corpus are shown, with a small subset of ingredients dominating usage. **g**, Distribution of cooking actions: Frequencies of common cooking techniques reveal a structured vocabulary of culinary operations, with a limited set of actions accounting for the majority of procedural steps. Together, this framework establishes a structured representation of global culinary knowledge enabling systematic analysis of ingredient usage, diversity, and compositional structure.*

**A global corpus of structured recipes**

Here we address this question by systematically analyzing the statistical properties of recipes at scale. We assemble a curated corpus of 118,083 traditional recipes[10] from diverse world cuisines, compiled from online sources and spanning 26 cuisines across 75 countries (Fig. 1a, Supplementary Fig. 1a, Supplementary Table 1, Supplementary Table 2). Each recipe was preprocessed and annotated using a named entity recognition (NER) pipeline[14] that identifies ingredients, preprocessing steps, cooking techniques, utensils, and other culinary descriptors[40] (Fig. 1b). Using manually annotated, augmented, and machine-annotated ingredient phrases, culinary entities were extracted with a fine-tuned spaCy-transformer NER model[14] (Supplementary Table 3 and Supplementary Fig. 2). This structured representation of recipes allows us to treat recipes as compositional objects built from multiple types of culinary units and to quantify their usage patterns across and within cuisines.

The resulting dataset contains 1.15 million mentions of 19,019 unique ingredients that are processed with 270 unique cooking techniques (Supplementary Table 4 and Supplementary Table 5). Recipes vary widely in their complexity, with an average of ten ingredients (Fig. 1c), an average of nine steps per recipe (Fig. 1d), and a median cook time of 30 minutes. A dish can thus be viewed as an emergent entity arising from the combination of ingredients and culinary processes with tasty and nutritious outcomes (Fig. 1e). The most frequently used ingredients include salt, onion, butter, flour, and oil (Fig. 1f). Frequently occurring cooking actions range from simple operations such as 'add' and 'place' to more specific transformations such as 'heat', 'cook', and 'bake' (Fig. 1g and Supplementary Table 5). Extended Data Fig. 1 and Extended Data Fig. 2 provide detailed statistics of these entities across the corpus. Having compiled the structured data of global recipes, we examined whether canonical laws arise in the culinary domain.

**Ingredient usage follows Zipf-like laws**

We first probed the distribution of ingredient usage frequencies by constructing rank-frequency plots for the global corpus. When ranked by decreasing frequency, ingredient counts follow an approximately linear relationship on log–log axes over several orders of magnitude (Fig. 2a), indicative of a Zipf-like distribution. Fitting a discrete power-law model[41] yields an exponent of $\alpha_{Global} \approx 1.53$, with alternative models such as log-normal ($p \sim 0.0084$) or exponential ($p \sim 10^{-4}$) providing significantly poorer fits.

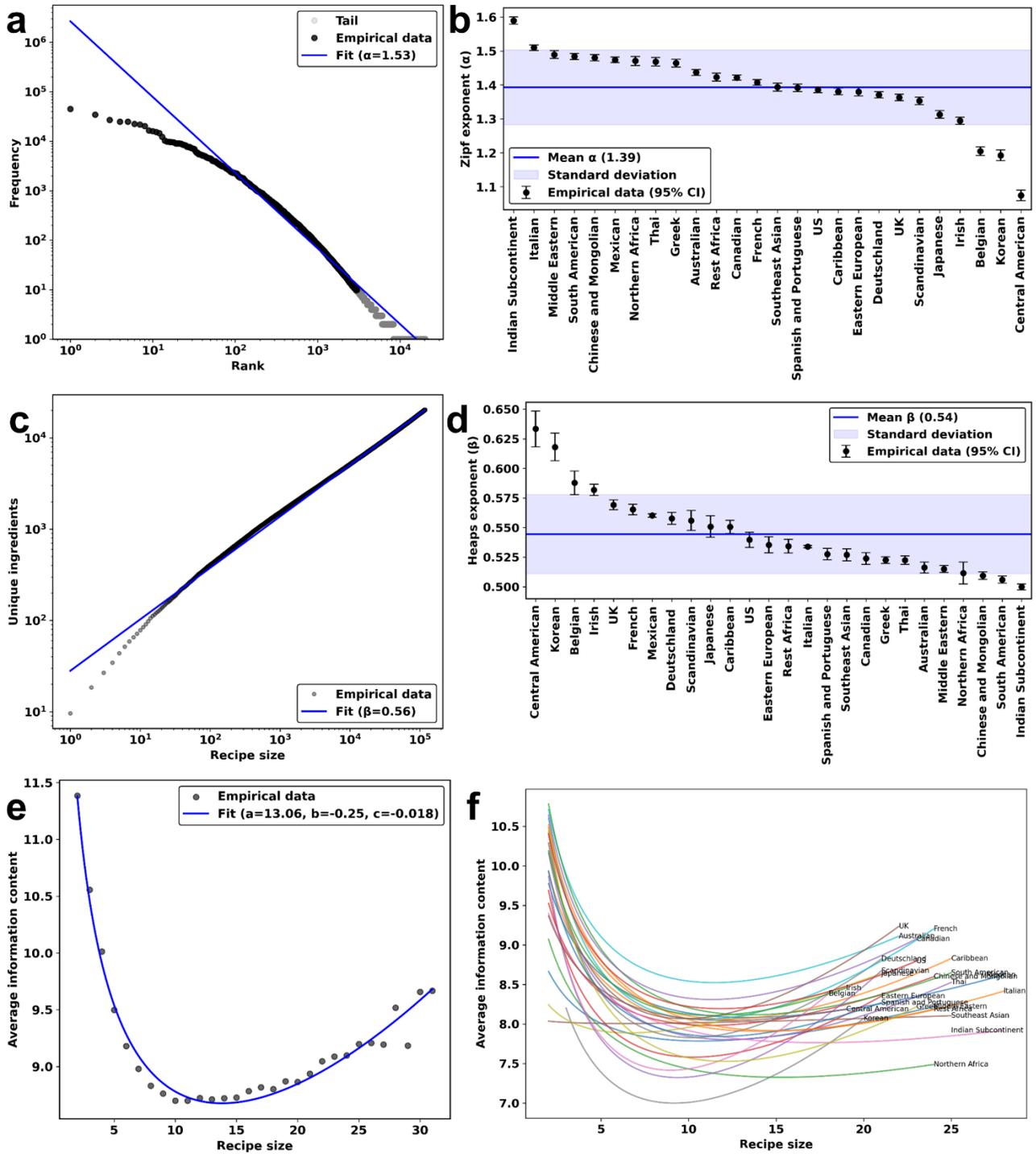

*Fig. 2 | **Statistical patterns governing culinary composition across global cuisines.** Large-scale analysis of ingredient usage and recipe structure reveals consistent scaling laws characteristic of complex compositional systems. **a,** Zipf-like rank–frequency distribution: Ingredient frequencies in the global corpus follow a heavy-tailed distribution when ranked by usage frequency. The linear trend in log–log space indicates Zipf-like scaling, $f(r) \sim r^{-\alpha}$, with exponent $\alpha \approx 1.53$. **b,** Cross-cuisine variability of Zipf exponent: Estimated Zipf exponents across 26 cuisines show a narrow distribution around the global mean ($\bar{\alpha} \approx 1.39$), with shaded regions indicating one standard deviation and error*

*bars representing 95% confidence intervals. Despite cultural and regional diversity, cuisines exhibit remarkably consistent scaling behavior. **c,** Heaps' law for culinary diversity: The growth of unique ingredients $V(R)$ with increasing number of recipes R follows sublinear scaling, $V(R) \sim R^\beta$, with $\beta \approx 0.56$. This indicates diminishing returns in ingredient diversity as more recipes are considered, consistent with constrained exploration of the ingredient space. **d,** Cross-cuisine variability of Heaps exponent: Heaps exponents across cuisines are tightly distributed around the global mean ($\bar{\beta} \approx 0.54$), again with limited dispersion across culinary traditions. **e,** Menzerath–Altmann law in recipe composition: The relationship between recipe size and average information content per recipe exhibits a non-linear trade-off consistent with the Menzerath–Altmann law. As recipe size increases, average information content initially decreases, reflecting the inclusion of more common ingredients, followed by a gradual increase at larger sizes. The fitted curve captures this balance between complexity and redundancy. **f,** Cuisine-specific Menzerath–Altmann trends: Individual cuisines display qualitatively similar trends, with variations in the position and curvature of the minima. This consistency across cuisines indicates a shared organizational principle governing the balance between recipe length and informational complexity. Collectively, these findings point to consistent statistical regularities across diverse culinary traditions.*

This Zipf-like behavior is robust across individual cuisines. For each of the 26 cuisines, we estimated the Zipf exponent from the ingredient frequency distribution. Exponents cluster tightly around a mean value of $\alpha \approx 1.392$, with 21 cuisines falling within one standard deviation of the mean (Fig. 2b and Supplementary Table 6). These results indicate that ingredient usage in recipes exhibits heavy-tailed scaling with a narrow range of exponents across diverse culinary traditions, consistent with a structured culinary 'vocabulary' shaped by preferential reuse. While the Zipf-like signature is consistently found across the cuisines, the exact set of dominant ingredients and those used infrequently in each cuisine characterizes its uniqueness. Incidentally, the popularity of ingredients is known to be a critical factor for rendering the food pairing patterns observed in cuisines[4–6].

**Culinary diversity grows sublinearly**
To quantify how ingredient diversity accumulates with increasing numbers of recipes, we analyzed the relationship between the number of distinct ingredients $V(R)$ and the number of recipes $R$, analogous to Heaps' law in language. For the global corpus, $V(R)$ grows sublinearly with $R$ on log–log scales, and is well described by a power-law relation $V(R) = R^\beta$ with $\beta_{Global} \approx 0.56$ (Fig. 2c). This indicates diminishing returns in ingredient discovery, consistent with vocabulary growth in symbolic systems—as more recipes are added, new ingredients continue to appear but at a decreasing rate.

The sublinear growth pattern generalizes across cuisines. Cuisine-specific Heaps exponents ($\beta_{Cuisine}$) fall within a relatively narrow band, with a mean of 0.5441 and 19 of 26 cuisines lying within one standard deviation of that mean (Fig. 2d and Supplementary Table 7). We also observe a systematic relationship between $\alpha$ and $\beta$: cuisines with steeper Zipf exponents tend to have smaller Heaps exponents (Extended Data Fig. 3), consistent with theoretical links between Zipf and Heaps laws in other symbolic systems[16,26]. Thus, Zipf and Heaps patterns are not independent laws in cooking; rather, they likely reflect a shared balance between reuse of common ingredients and the introduction of novel ones. The more a cuisine relies on a few dominant ingredients, the slower it introduces new ingredients.

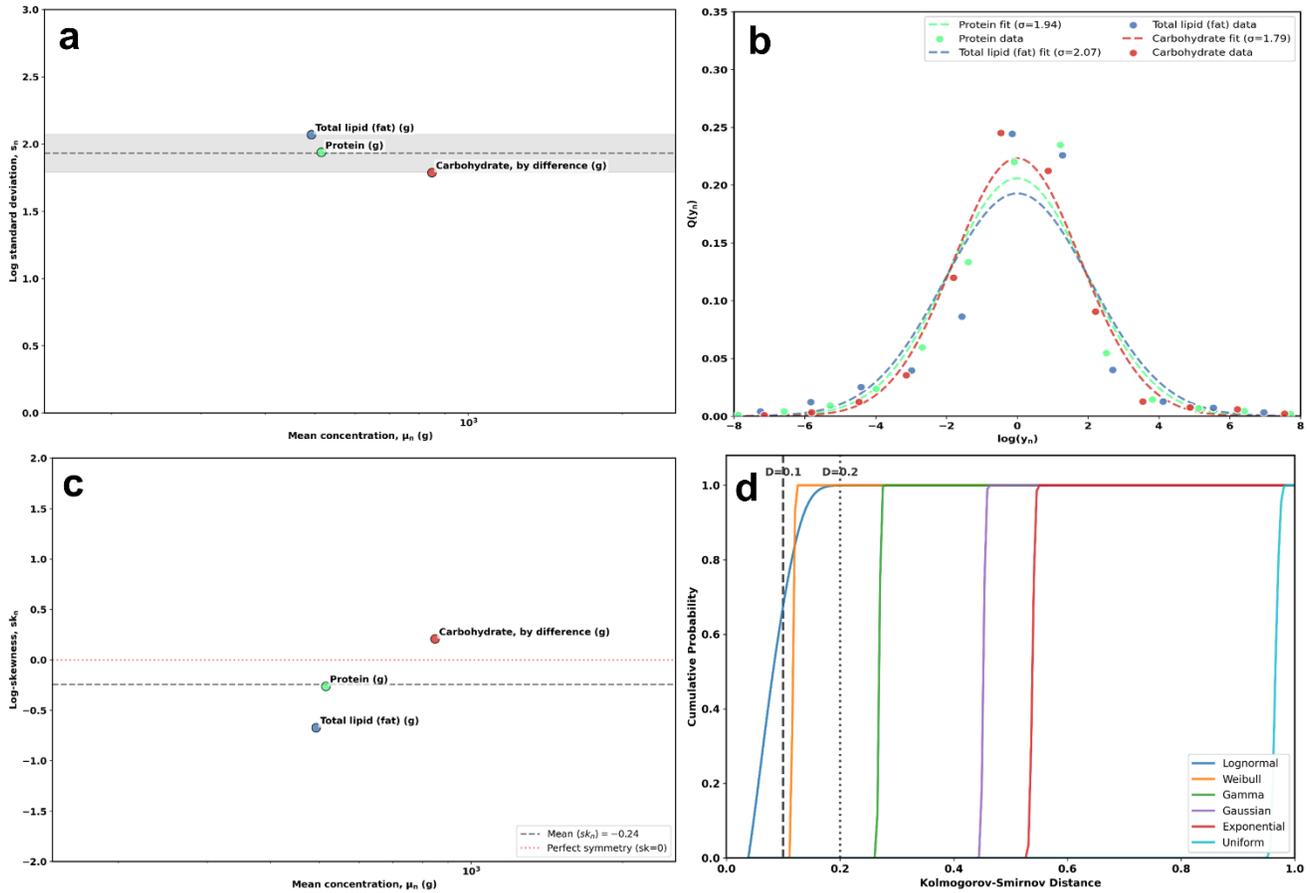

*Fig. 3 | Statistical signatures in macronutrient composition across global recipes. Macronutrient concentrations—carbohydrates, proteins, and total lipids (fats)—exhibit consistent statistical regularities across a large corpus of recipes spanning diverse cuisines. **a**, Mean–variance relationship: The logarithm of the standard deviation ($s_n$) of macronutrient concentrations is plotted against the mean concentration ($\mu_n$) across recipes. All three macronutrients lie within a narrow band of variability, indicating a constrained dispersion, independent of scale. The shaded region highlights the limited range of fluctuations, suggesting a common underlying generative constraint governing nutrient distributions (Also see Extended Data Fig. 6a). **b**, Translational invariance of distributions: Probability density functions of the logarithm of macronutrient concentrations, $\log(y_n)$, collapse onto approximately symmetric, unimodal curves for all three macronutrients. The close overlap between empirical distributions and fitted curves is consistent with invariance under translation in log-space, a hallmark of log-normal behavior. **c**, Symmetry of distributions: Log-skewness ($s_{k_n}$) of macronutrient concentrations is close to zero across nutrients, indicating approximate symmetry in log-space. Deviations from perfect symmetry are small and bounded, further supporting the hypothesis of log-normality (Also see Extended Data Fig. 6b). The dashed line denotes the mean skewness across nutrients, while the dotted red line represents perfect symmetry ($s_k = 0$). **d**, Goodness-of-fit analysis: Kolmogorov–Smirnov statistics comparing empirical distributions with candidate theoretical*

*distributions demonstrate that the log-normal distribution consistently provides the best fit (lowest KS distance). Vertical reference lines indicate representative KS thresholds, highlighting the clear separation between log-normal fits and alternative models. Together, these observations provide evidence that macronutrient compositions in global recipes follow a log-normal distribution. This statistical regularity suggests the presence of underlying processes and constraints in culinary design. Detailed analyses for individual macronutrients are presented in Extended Data Fig. 5, and cuisine-specific statistics for the ten largest cuisines are provided in Extended Data Fig. 6.*

**Complexity reduces with increasing recipe size**

Beyond frequency and diversity, we asked whether recipes exhibit trade-offs between the number of constituent units and their average complexity, in line with the Menzerath–Altmann law[19,20]. We defined 'recipe size' as the number of ingredients, and 'ingredient complexity' as an information-theoretic measure based on ingredient rarity. The rarer an ingredient is in the corpus, the higher its complexity, or information content. Across the corpus, recipes with more ingredients tend to have lower average ingredient complexity, whereas shorter recipes tend to use on average more informative ingredients (Fig. 2e). The relationship is well captured by a Menzerath-Altmann function of the form $y(L) = a \cdot L^b \cdot e^{cL}$, with parameters $(a, b, c)$ that vary modestly across cuisines (Fig. 2f, Supplementary Fig. 3, and Supplementary Table 8). The non-monotonic dependence on size reveals a fundamental trade-off between efficiency and expressivity in culinary design.

Similar statistical regularities are observed when considering an alternative unit, namely cooking techniques used in recipes (Extended Data Fig. 4). These patterns suggest a general design principle in which increasing the number of units within a recipe is offset by reduced average complexity of those units, potentially reflecting cognitive and practical constraints on what can be managed in the process of cooking.

**Log-normality in nutrient concentrations**

Careful composition of ingredients and their transformation through cooking yields nutritional dishes (Fig. 1e and Supplementary Fig. 1b). The nature of patterns in nutritional content of food has been a longstanding question[42,43]. We therefore asked whether, analogous to packaged food products[15], macronutrient concentrations in cooked meals exhibit a consistent statistical signature across global cuisines. Remarkably, macronutrient concentrations in cooked recipes follow a log-normal distribution across culinary systems (Fig. 3, Extended Data Fig. 5, Extended Data Fig. 6, Supplementary Figs. 4–8). Carbohydrates, fats, and proteins exhibit narrow dispersion in log-space (Fig. 3a), approximate translational invariance after rescaling (Fig. 3b), and near-symmetry in log-space as quantified by logarithmic skewness (Fig. 3c). These signatures of log-normality are further supported by Kolmogorov–Smirnov statistics, for which the log-normal distribution provides the best fit among the candidate models tested (mean K–S distance: log-normal, 0.0727; Weibull, 0.1171; Gamma, 0.2687; Gaussian, 0.4520; Exponential, 0.5383; Uniform, 0.9662) (Fig. 3d). Together, these results indicate that nutrient concentrations display a consistent multiplicative statistical structure across diverse culinary traditions (Also see Supplementary Figs. 4—8). While Zipf and Heaps laws govern the combinatorial organization of recipes, the log-normal distribution of nutritional attributes is consistent with multiplicative aggregation across culinary components.

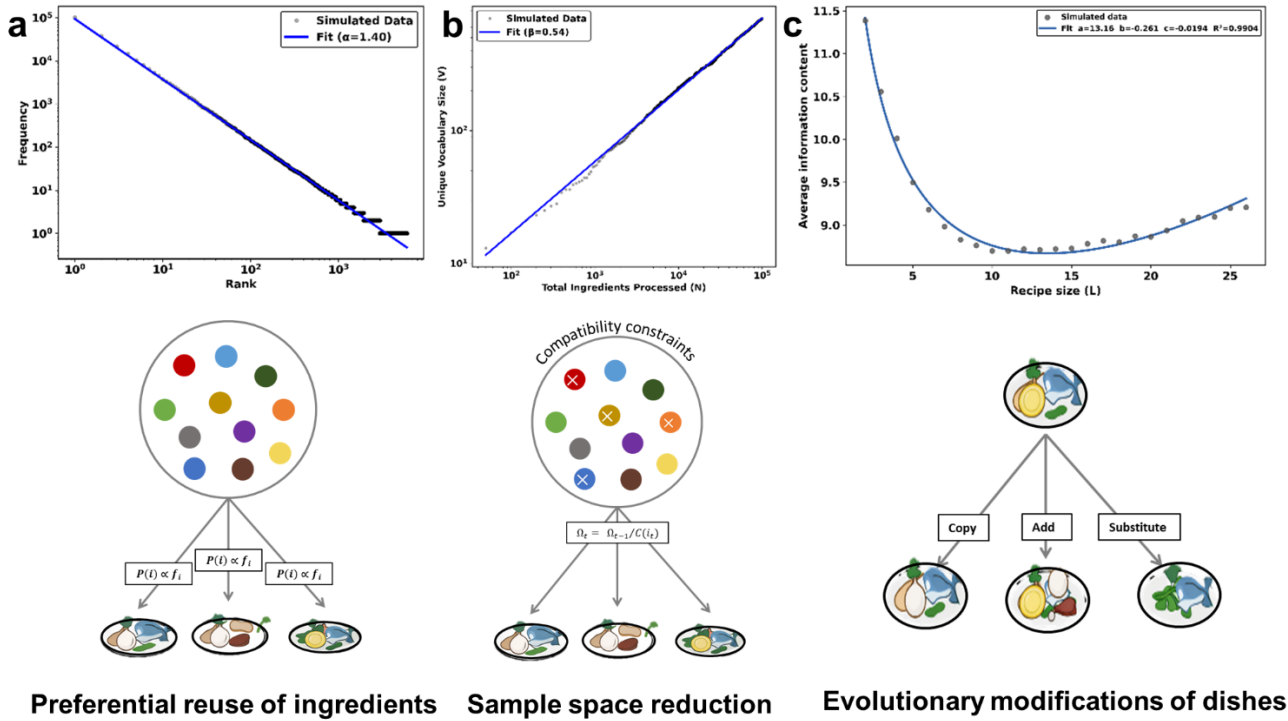

***Fig. 4 | Generative models reproducing culinary laws.*** *Schematic representation of the minimal generative mechanisms underlying the emergence of statistical regularities in recipe composition. **a,** Preferential reuse (rich-get-richer): New recipes are constructed by sampling ingredients from an evolving ingredient pool with probability proportional to their historical frequency, $P(i) \propto f_i$. This biased selection mechanism amplifies the usage of already popular ingredients, leading to heavy-tailed rank–frequency distributions consistent with Zipf's law. **b,** Constraint-driven sample-space reduction: Ingredient selection is conditioned on compatibility constraints arising from flavor, functional, cultural, or preparation-based affinities, such that the effective sampling probability becomes $P((i|\Omega_t))$, where $\Omega_t$ denotes the context defined by ingredients already present in the recipe. This progressive restriction of the accessible ingredient space limits combinatorial explosion and yields sublinear vocabulary growth consistent with Heaps' law. **c,** Evolutionary modification of recipes: New recipes are generated through incremental transformations of existing ones via addition, removal, or substitution of ingredients, capturing processes of culinary innovation and adaptation. These evolutionary dynamics introduce a trade-off between recipe length and informational content, giving rise to Menzerath–Altmann-type relationships between the number of ingredients and their average information. Together, these mechanisms illustrate a minimal generative framework capable of reproducing the observed scaling laws in culinary data. Preferential reuse drives frequency heterogeneity, constraint-based selection shapes the accessible combinatorial space, and evolutionary modifications govern the structural organization of recipes. The interplay of these processes demonstrates how simple local rules can give rise to statistical patterns in a complex, culturally embedded system.*

**Generative models reproduce culinary laws**

These results suggest that ingredient usage in recipes exhibits heavy-tailed scaling, with a narrow range of exponents across diverse culinary traditions. To probe whether simple stochastic mechanisms can account for these regularities, we implemented minimal generative models for the principal empirical patterns. The fitness-driven preferential sampling model yields a heavy-tailed rank–frequency distribution with exponent $\alpha = 1.40$, closely matching the empirical Zipf scaling. This result suggests that a simple rank-based preferential sampling mechanism is sufficient to generate the observed heterogeneity in ingredient usage. The Zipfian sampling model for vocabulary growth produces sublinear scaling of ingredient diversity with corpus size, with an estimated Heaps exponent $\beta = 0.54$. This behavior reflects diminishing returns in ingredient discovery and arises naturally from repeated sampling over a heavy-tailed ingredient distribution, consistent with constrained exploration in an effective sense.

To capture the Menzerath–Altmann trade-off, we fitted the dependence of average ingredient complexity on recipe length. The global fit yields parameters $a = 13.16, b = -0.26, and\ c = -0.019$ with high goodness-of-fit ($R^2 = 0.99$), and exhibits a characteristic minimum at intermediate recipe lengths ($L \approx 13$). This non-monotonic behavior indicates that longer recipes tend to incorporate more common ingredients, balancing expressivity with compositional efficiency. Taken together, these results show that simple stochastic models grounded in preferential sampling and compositional constraints can reproduce the principal scaling laws of culinary design (Fig. 4).

**Discussion**

Human culinary practice spans geography, climate, history, and cultural identity; yet our results reveal that recipes across the world exhibit consistent statistical patterns. Ingredient frequencies follow Zipf-like heavy-tailed scaling, the accumulation of novel ingredients is sublinear in accordance with Heaps' law, and recipe structure obeys Menzerath–Altmann-type trade-offs between recipe size and ingredient complexity. These empirical laws mirror those observed in natural languages, cities, and other symbolic systems, suggesting that cooking is not only a cultural activity but also a complex generative process governed by universal constraints and amplification dynamics. Taken together, these findings suggest that recipes can be understood as compositional symbolic systems whose large-scale structure is shaped by statistical constraints.

The emergence of Zipf and Heaps laws in culinary data points to a fundamental balance between reuse and novelty. A small number of core ingredients dominate global recipes, while a long tail of rare ingredients sustains diversity. Sublinear diversity growth indicates that, as cuisines expand, novelty persists yet slows—echoing vocabulary growth in languages and diversification in biological and technological systems. Together, these patterns suggest that culinary knowledge evolves through a constrained exploration of ingredient space, where repeated reuse confers stability and discoverability, and incremental innovation introduces diversity.

Minimal models incorporating preferential reuse, constrained sampling, and incremental modification recapitulate the observed scaling laws at a coarse-grained level. Preferential sampling reproduces heavy-tailed ingredient frequency distributions, and Zipfian vocabulary-growth models account for the observed sublinear expansion of ingredient diversity. The Menzerath–Altmann relation further captures the trade-off between recipe length and average ingredient complexity. Although these models are intentionally minimal, their ability to reproduce the main empirical regularities suggests that culinary structure can arise from simple generative and compositional principles rather than culturally specific rules alone.

Previous studies on ingredient co-occurrence and flavor pairing have revealed strong cuisine-specific patterns, such as positive[3] or negative pairing[4,5] tendencies arising from shared flavor compounds. In contrast, the statistical regularities identified here—Zipf, Heaps, and Menzerath–Altmann laws—operate at a higher level of abstraction, capturing universal constraints on ingredient usage, diversity, and compositional structure. Notably, cuisine-specific pairing biases can be interpreted as modulating the local organization of ingredients within recipes, while still being embedded within global scaling laws that govern the overall architecture of culinary systems.

These findings position cooking within the broader class of complex systems in which creativity arises from simple rules repeated at scale. While cuisine remains a deeply human expression of identity and culture, our results indicate that this creativity is scaffolded by universal statistical principles. In doing so, we provide a quantitative foundation for computational gastronomy as a scientific discipline[38,39], linking culinary practice to established theories of complex and compositional systems.

Beyond cultural insight, the presence of statistical culinary laws opens avenues for systematic exploration of culinary processes and recipe design. Statistical models of recipe structure could guide the synthesis of new dishes, enable nutrition-aware reformulations, and identify minimal interventions required to improve dietary profiles while preserving cultural integrity. More broadly, understanding the mathematical structure of recipes offers a principled framework for navigating the vast space of possible dishes—towards novel combinations, improved nutrition, and scalable food innovation.

Our findings, however, should be interpreted in light of several limitations. The corpus, assembled from online repositories, may reflect reporting and availability biases, including uneven representation across cuisines and sources. Cuisine labels, derived from metadata and manual curation, provide a useful but imperfect proxy for complex and evolving culinary identities. The observed patterns are also contingent on ingredient normalization and structured annotation, which may introduce aggregation effects in ingredient usage and categorization. Finally, these observations establish robust phenomenological regularities, but do not uniquely identify the underlying generative mechanism.

Future work can extend this framework by incorporating factors such as ingredient taxonomy, sensory chemistry, cost, climate, and supply chains. Integrating flavor molecule networks and metabolic responses may reveal deeper constraints shaping recipe structure, while historical and temporal data can help trace how these patterns emerge and evolve over time. More broadly, our findings place recipes within the wider class of symbolic systems whose diversity is shaped by universal statistical constraints, opening a quantitative route to understanding, comparing, and ultimately designing culinary structure.

## Methods

**Dataset collection and consolidation**

We assembled a corpus of recipes representing culinary traditions across the world[10] (Fig. 1). Recipes were aggregated from a wide variety of public online repositories[10]. Each document contained natural language descriptions of ingredient phrases comprising ingredients, quantities, their preprocessing attributes, and instructions describing processing steps, techniques, and utensils. The dataset was harmonised by removing duplicate entries, normalising encoding formats, and correcting structural irregularities such as misformatted ingredient lists.

Regional attribution was assigned using source metadata wherever available; cuisine labels were manually curated in case of ambiguities. Only recipes with sufficient annotation completeness were retained. The final corpus contained 118,083 recipes from 75 countries spanning 26 cuisine types (Fig.1a and Supplementary Fig. 1a). The statistics of recipes from across cuisines and the details of the cuisines schema are provided in Supplementary Tables 1 and Supplementary Table 2, respectively.

**Named entity recognition and recipe structuring**

To convert free-text recipes into structured representations, we developed an entity extraction pipeline[12–14] grounded in named entity recognition (NER). Models were trained to identify and classify principal culinary entity groups: ingredients, quantity, unit, preprocessing operations (e.g., chopped, roasted, marinated), cooking techniques (e.g., fry, steam, sauté), utensils/equipment (e.g., skillet, tandoor, wok) and more.

We created an augmented dataset of 26,446 phrases by building on manually annotated sets comprising 6,611 ingredient phrases (training) and 2,187 labeled phrases (test) compiled from the global recipes corpus. We systematically cleaned and analyzed ingredient phrases and annotated them using Stanford NER. We sampled a subset of 88,526 phrases using a clustering-based approach while preserving the diversity to create the machine-annotated dataset. A thorough investigation of NER approaches across these datasets involving fine-tuning of deep learning-based language models and few-shot prompting on large language models led to the construction of a state-of-the-art fine-tuned spaCy-transformer model with macro-F1 scores of 95.9%, 96.04%, and 95.71% for the manually-annotated, augmented, and machine-annotated datasets, respectively (Supplementary Table 3)[14]. Culinary entities were recognised using this spaCy-transformer model. The summary statistics for various named entities are shown in Extended Data Fig. 1 and Extended Data Fig. 2.

**Quantifying Zipf's law of ingredient frequency**

We quantified ingredient usage by computing the frequency $f_i$ of each unique ingredient $i$, defined as the number of recipes in which the ingredient appears. Ingredients were ranked in descending order of frequency to obtain rank $r$, and the resulting rank–frequency distribution $f(r)$ was examined on logarithmic axes. To rigorously test for power-law scaling, we fitted a discrete power-law model of the form $P(f) \sim f^{-\alpha}$ using maximum-likelihood estimation following the framework of Clauset *et al*[41,44]. The lower bound $f_{min}$ was determined by minimizing the Kolmogorov–Smirnov (K–S) distance between the empirical and fitted distributions. The Zipf exponent $\alpha$ was estimated for the global corpus and for individual cuisines (Fig. 2a and Fig. 2b). Model validity was assessed via likelihood ratio tests comparing power-law fits against log-normal, and exponential alternatives. Goodness-of-fit was quantified using the K–S statistic and associated $p$-values. Only fits with $p > 0.1$ were considered statistically plausible.

**Heaps' law and culinary diversity growth**
To characterize the growth of ingredient diversity with corpus size, we measured the number of distinct ingredients $V(R)$ observed as a function of the number of recipes $R$. We performed incremental sampling experiments by randomly shuffling recipes 100 independent times. For each permutation, we computed the cumulative number of unique ingredients $V(R)$. Heaps' law was fitted using $V(R) = KR^\beta$, where $\beta$ is the Heaps exponent and $K$ is a normalization constant. Exponent $\beta$ was estimated via least-squares regression on log–log transformed data. Error bounds were obtained from variability across the 10 random permutations. To test universality, we performed data-collapse analysis using the transformation $V(R)/R^\beta \sim constant$. A successful collapse across cuisines indicates shared scaling behavior. Cuisine-specific exponents were computed for the global corpus and for each cuisine separately (Fig. 2c, Fig. 2d, and Supplementary Table 7).

**Menzerath–Altmann law in recipe complexity**
We investigated structural trade-offs in recipe composition using the Menzerath–Altmann framework. Recipe length was defined as the number of unique ingredients in a recipe ($L$). Ingredient complexity was defined as $s_i = -\log p(i)$, where $p(i) = f_i/\sum_i f_i$ is the empirical probability of ingredient $i$. Average recipe complexity was computed as $y(L) = \frac{1}{L}\sum_{i=1}^{L} s_i$. This definition captures the intuition that rarer ingredients carry higher informational complexity. We fitted the Menzerath–Altmann relation $y(L) = a \cdot L^b \cdot e^{cL}$, where $a$, $b$, and $c$ are free parameters. Parameter estimation was performed using nonlinear least-squares regression. Model quality was evaluated using coefficient of determination ($R^2$). Fits were computed for the global dataset and individual cuisines (Fig. 2e, Fig. 2f, and Supplementary Table 8). To test robustness, we repeated the analysis with alternative definitions of units as cooking techniques instead of ingredients.

**Nutrient Analysis**
**Nutrient prevalence analysis.** The macronutrients data (carbohydrate, protein, and fat) were obtained from the global culinary corpus[10] along with cuisine labels of recipes. The dataset contains 118,083 recipes spanning 26 cuisines. Nutrient values were standardized, and per-serving data were used for analysis.

As a preliminary characterization step, we computed the prevalence $p_n$ of each macronutrient, defined as the fraction of recipes in which that nutrient is present at a non-zero level: $p_n = \frac{Number\ of\ recipes\ with\ x_n > 0}{Total\ number\ of\ recipes}$, where $x_n$ is the amount of a nutrient '$n$' (in grams) recorded for a given recipe. This metric quantifies how commonly each macronutrient appears across recipes within a given cuisine. Prevalence values were computed separately for the complete dataset and for each of the cuisines, providing a baseline measure of macronutrient representation prior to distributional analysis.

**Distribution of nutrient content $Q(x_n)$.** To characterize variation of nutrient content across recipes, we computed the probability distribution $Q(x_n)$ for each macronutrient, where $x_n$ is the amount (in grams) of nutrient $n$ in a recipe and $Q(x_n)$ denotes the normalized probability density of observing $x_n$. For each nutrient, we applied a log transformation to all positive values: $z_n = \ln(x_n)$. Normalised histograms of the log-transformed values were constructed using 15 equal-width bins. The bin centres were back-transformed to the original scale: $x_{center} = \exp\left(\frac{z_{left}+z_{right}}{2}\right)$, where $z_{left}$ and $z_{right}$ represent the left and right edges (in log-space) of a given histogram bin, respectively. A Gaussian distribution was then fitted to each log-transformed dataset with maximum likelihood estimation, yielding parameters $\mu$ (log-mean) and $\sigma$ (log-standard deviation) (Extended Data Fig. 6 and Supplementary Fig. 4). A Gaussian shape in the log-space is the hallmark of a log-normal distribution in the original linear scale. The fitted curve represents: $Q(x_n) \sim Log-Normal(\mu, \sigma)$.

**Constant standard deviation in log space ($s_n$).** A central expectation under the log-normal hypothesis is that the spread of nutrient distributions, when measured in logarithmic space, remains approximately constant regardless of the mean nutrient level. To test this, we computed the log-standard deviation $s_n$ for each nutrient: $s_n = \sqrt{\frac{1}{N_n-1}\sum_{r=1}^{N_n}\left(\ln\left(x_n^{(r)}\right) - m_n\right)^2}$, where $x_n^{(r)}$ is the nutrient content of recipe $r$, $m_n = \frac{1}{N_n}\sum_r \ln\left(x_n^{(r)}\right)$ is the log-mean, and $N_n$ is the number of recipes with non-zero nutrient values. The log-standard deviation $s_n$ quantifies the spread of each nutrient's distribution in log-space. The values of $s_n$ were plotted against the arithmetic mean $\mu_n$ for all three macronutrients. The mean and standard deviation across all nutrients, $\langle s_n \rangle \pm \sigma_{s_n}$, were computed and displayed as a shaded reference band (Fig. 3a and Extended Data Fig. 6a), where $\langle s_n \rangle$ is the mean of all log-standard deviation values and $\sigma_{s_n}$ is their standard deviation. A band spanning $\langle s_n \rangle - \sigma_{s_n}$ to $\langle s_n \rangle + \sigma_{s_n}$ (i.e. mean $\pm$ one standard deviation) was overlaid on the plot as a visual reference for assessing the constancy of $s_n$ across nutrients.

**Symmetry analysis (Logarithmic skewness ($sk_n$)).** Log-normality further requires that distributions be symmetric in log-space. To verify this property, we computed the logarithmic skewness $sk_n$ for each macronutrient: $sk_n = \frac{1}{N_n}\sum_{r=1}^{N_n}\left(\frac{\ln\left(x_n^{(r)}\right)-m_n}{s_n}\right)^3$. The logarithmic skewness $sk_n$ measures the asymmetry in each nutrient's distribution in the log-space and is defined as the standardized third central moment of the log-transformed values. Skewness values close to zero indicate approximate symmetry (Gaussian-like behaviour in the log-space), while positive values indicate a heavy upper tail and negative values indicate a heavy lower tail. We plotted $sk_n$ against the mean $\mu_n$ for each nutrient and found a mean skewness of $\langle sk_n \rangle = -0.24$, indicating an overall slight left-skew in log-space (Fig. 3c and Extended Data Fig. 6b). Carbohydrate showed mild positive skew, while protein and fat were negative, with values remaining relatively close to the symmetry line at $sk_n = 0$.

**Translational invariance—rescaled distribution $Q(y_n)$.** Translational invariance in log-space—a defining property of log-normal distributions—implies that distributions for different nutrients should collapse onto a single universal curve once each is centred on its log-mean. To test this, we applied the rescaling transformation: $y_n = \exp(\ln(x_n) - m_n) \Leftrightarrow \ln(y_n) = \ln(x_n) - m_n$. Here, $y_n$ is the rescaled nutrient variable obtained by centering $\ln(x_n)$ around its mean $m_n$. This operation shifts each distribution so that its log-mean is centred at zero. We computed $Q(y_n)$ for carbohydrate, fat, and protein, and assessed the degree of collapse onto a common curve. A Gaussian was fitted to each rescaled distribution to obtain comparable spread parameters, providing a quantitative summary of the universality hypothesis (Fig. 3b and Supplementary Fig. 8). The resulting spread parameters were: $\sigma \approx 1.79$ (carbohydrate), $\sigma \approx 1.94$ (protein), and $\sigma \approx 2.07$ (fat).

**Power-law scaling ($\sigma_n$ *versus* $\mu_n$).** We further investigated whether the standard deviation $\sigma_n$ scales with the arithmetic mean $\mu_n$ according to a power law—a relationship known as Taylor's law—as predicted by the log-normal model. The relevant quantities are defined as: $\mu_n = \frac{1}{N_n}\sum_{r=1}^{N_n} x_n^{(r)}$ and $\sigma_n = \sqrt{\frac{1}{N_n-1}\sum_{r=1}^{N_n}\left(x_n^{(r)} - \mu_n\right)^2}$. Taylor's law predicts the power-law relationship: $\sigma_n \propto \mu_n^\beta$, which, on a log-log scale appears as a linear relationship: $\ln(\sigma_n) = \gamma \cdot \ln(\mu_n) + \ln(A)$. We performed linear regression of $\ln(\sigma_n)$ against $\ln(\mu_n)$ for all three macronutrients. The slope $\gamma$ estimates the scaling exponent and $R^2$ quantifies the goodness of fit. A slope close to unity ($\gamma \approx 1$) is the expected signature of log-normal distributions, where the coefficient of variation remains constant. We obtained a slope of $\gamma = 0.9009$ and $R^2 = 0.9656$, suggesting a near-linear scaling relationship consistent with log-normal behaviour (Supplementary Fig. 5).

**Kolmogorov–Smirnov goodness-of-fit test.** To rigorously identify the theoretical distribution that best describes the observed nutrient data, we applied the Kolmogorov–Smirnov (K–S) test against six candidate distributions (log-normal, Gamma, Weibull, Gaussian, Exponential, and Uniform) fitted to each macronutrient. For each nutrient–distribution pair, the K–S distance $D$ was computed as the maximum absolute difference between the empirical cumulative distribution function (CDF) and the fitted theoretical CDF: $D = \max_x |F_{empirical}(x) - F_{fitted}(x)|$. Smaller values of $D$ indicate a better fit. A value $D < 0.1$ is generally considered indicative of an excellent fit; values below 0.2 indicate an acceptable fit. The distribution achieving the lowest mean $D$ across all three macronutrients was identified as the best-fitting model (Fig. 3d and Supplementary Fig. 6).

**Generative model construction**
To probe the generative origins of the observed statistical regularities, we implemented simplified stochastic models that capture key aspects of ingredient reuse, constrained exploration, and compositional trade-offs. All simulations were performed on a synthetic ingredient space of size $V = 15{,}000$, consistent with the scale of the empirical corpus.

**Preferential reuse model.** To model Zipf-like frequency distributions, we employed a fitness-driven sampling process in which each ingredient $i$ was assigned a rank-dependent fitness $f_i \propto r_i^{-\alpha}$, with target global exponent $\alpha = 1.42$. Ingredient tokens were then sampled independently according to this fixed distribution for a total of $3 \times 10^5$ draws. The resulting rank–frequency distribution was fitted over the range $10 \leq r \leq 1000$ using maximum-likelihood estimation, yielding an emergent exponent consistent with empirical observations.

**Sample-space reduction model.** To reproduce Heaps' law, we simulated vocabulary growth using a Zipfian sampling framework. Ingredients were drawn from a master distribution with exponent $\alpha = 1.85$, selected to produce sublinear vocabulary scaling. A total of $10^5$ draws were generated, and the cumulative number of unique ingredients $V(R)$ was recorded at intervals of 50 tokens. The Heaps exponent $\beta$ was estimated via least-squares regression on log–log transformed data, excluding the initial transient regime ($R > 500$).

**Menzerath–Altmann structural model.** To quantify Menzerath–Altmann scaling, we analyzed the relationship between recipe length $L$ and average ingredient information content $y(L)$, defined using empirical ingredient probabilities. The functional form $y(L) = aL^b e^{cL}$ was fitted using nonlinear least-squares optimization with initial parameter estimates $(a, b, c)$=(13.0,−0.25,−0.018), bounded such that $b \leq 0$ and $c \leq 0$. Fits were performed both globally and across cuisines, requiring a minimum of 20 recipes per length bin and at least 5 distinct length values. Convergence was ensured with a maximum of 20,000 function evaluations.

## Code availability
The relevant data and code for the data analysis, statistical fitting, model simulation and figure generation are available at: https://github.com/cosylabiiit/Universal-statistical-laws-governing-culinary-design

## Acknowledgments
GB thanks Indraprastha Institute of Information Technology Delhi (IIIT-Delhi) for the computational support. This study was supported by the Infosys Center for Artificial Intelligence and Centre of Excellence in Healthcare at IIIT-Delhi. The authors thank Aryan Dahiya and Abhishek Rao for help in generating the global cuisine map. GB thanks Deepak Dhar, Ramakrishna Ramaswamy, and Sutirth Dey for comments and suggestions.

## Author Contributions
GB conceived the idea, designed the methodology, supervised the project, and wrote the manuscript. GKT, ARY, AS, PB, UD, MG, MKS, GB conducted data analysis and implemented models. All authors read and approved the final manuscript.

## Competing Interests
The authors declare no competing interests.

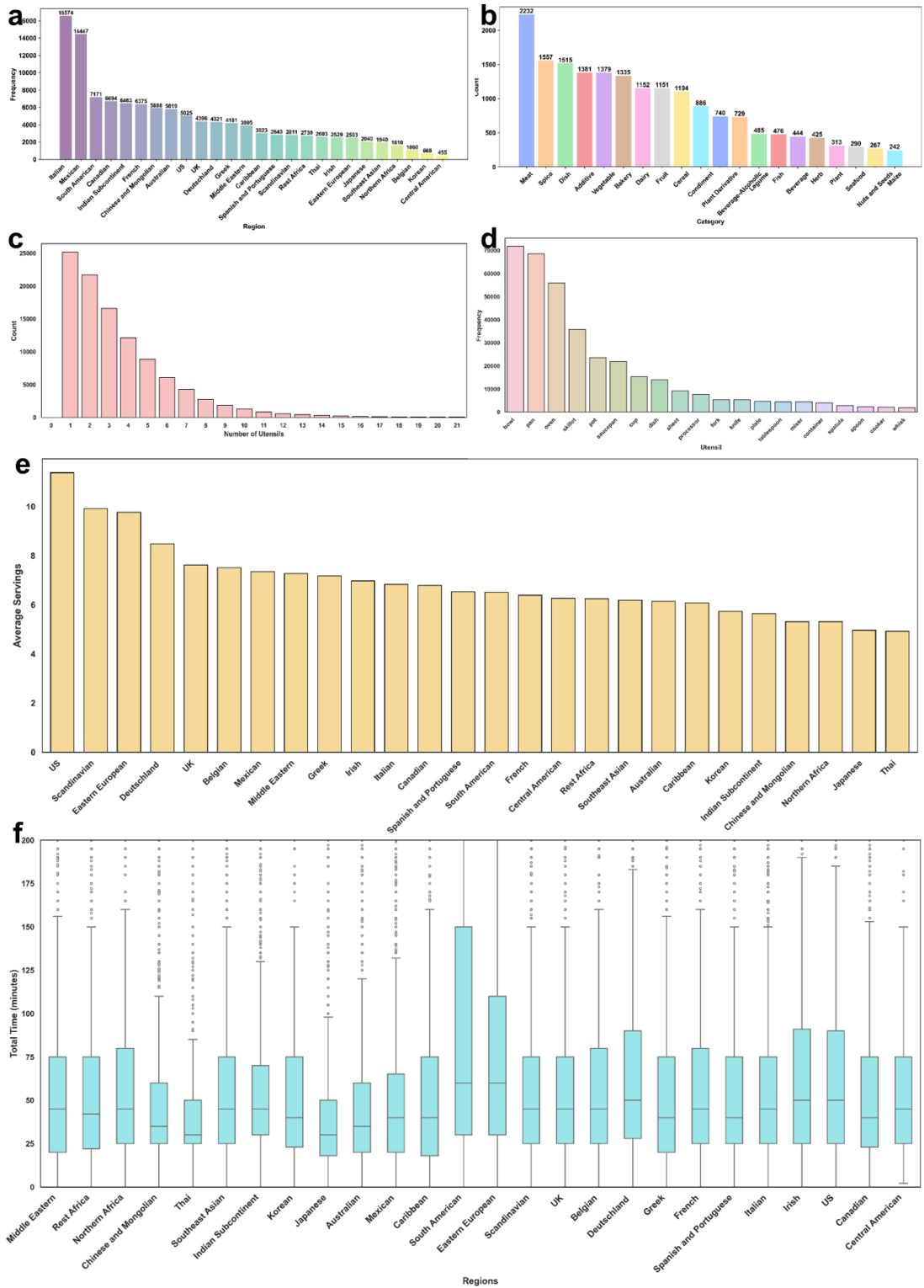

**Extended Data Fig. 1 | Comprehensive characterization of the global culinary corpus.** Summary statistics describing the composition, structure, and procedural attributes of the curated dataset. **a,** Distribution of recipes across cuisines: The number of recipes varies across cuisines, with larger representations for widely documented culinary traditions (e.g., Italian, Mexican, Indian Subcontinent, and South American) and smaller yet significant representations for others (e.g., Central American,

Korean, and Belgian). Major culinary traditions are sufficiently sampled for comparative analysis. **b,** Ingredient category distribution: Ingredients are grouped into 34 major categories. A few categories (e.g., meats, spices, vegetables) dominate, while less frequent categories contribute to long-tail diversity in ingredient usage. **c,** Distribution of utensils per recipe: The number of utensils used per recipe follows a rapidly decaying distribution, with most recipes requiring only a small set (1–5 utensils). This indicates that culinary processes are generally constrained in terms of operational complexity. **d,** Frequency of utensil usage: A small set of utensils accounts for most usage, with specialized tools appearing infrequently. **e,** Average servings per recipe: The mean number of servings per recipe varies moderately across cuisines, generally ranging between 5 and 11 servings. **f,** Distribution of total cooking time: Total preparation and cooking times exhibit broad distributions across cuisines, with median values typically between 20 and 60 minutes. Box plots illustrate substantial variability within cuisines, including long tails corresponding to time-intensive dishes. Despite this variability, the overall distributions remain comparable across cuisines, suggesting common temporal constraints in culinary practice. Together, these statistics provide a comprehensive overview of the dataset, highlighting both diversity and shared structural constraints across global culinary systems. This characterization establishes the robustness and representativeness of the corpus for uncovering statistical patterns in recipe composition.

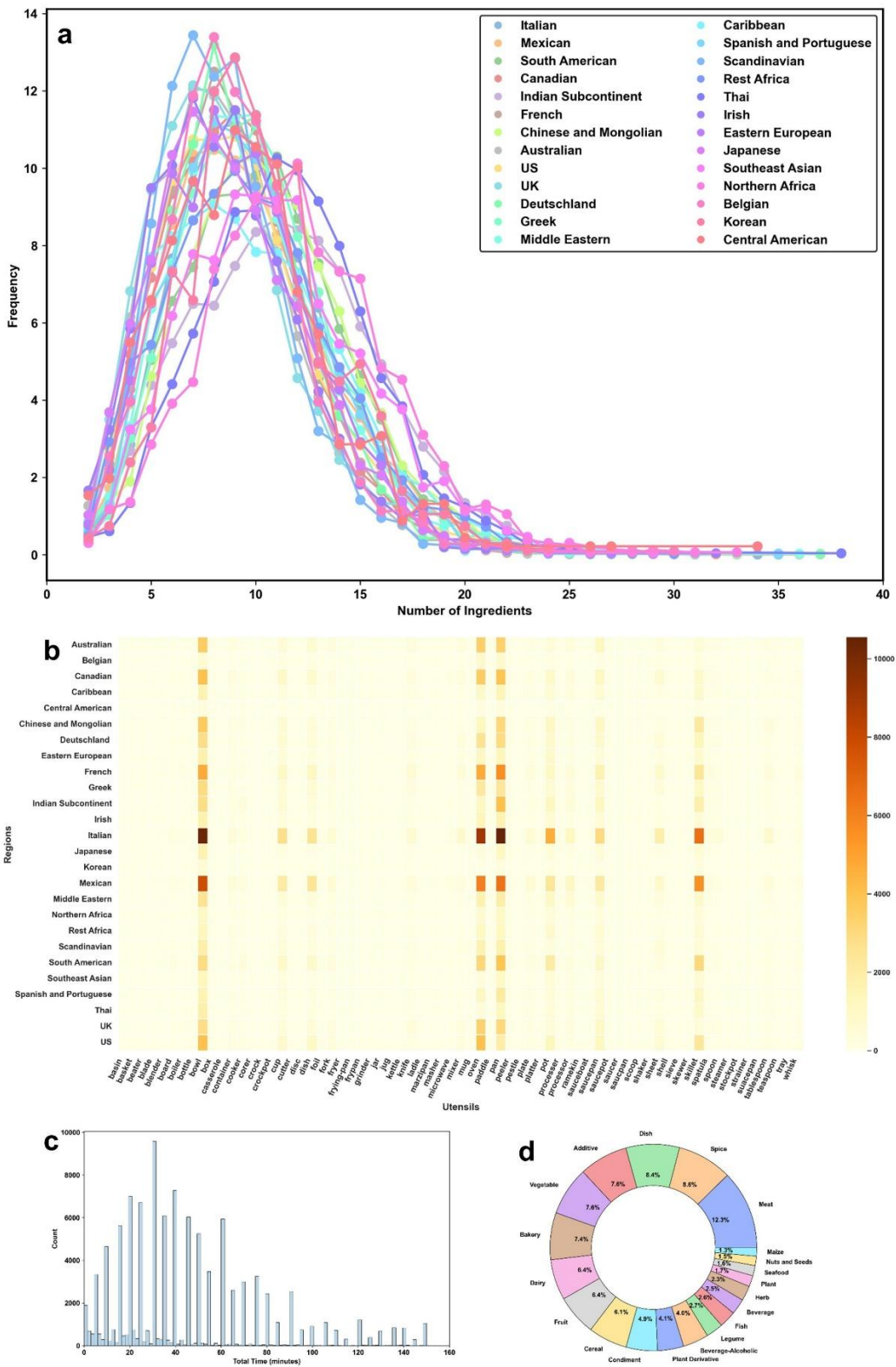

**Extended Data Fig. 2 | Statistical structure of the global, structured recipe corpus across cuisines.** Characterization of recipe composition, procedural complexity, and ingredient organization across 26 cuisines. **a,** Recipe size distributions across cuisines: The number of ingredients per recipe exhibits a consistent unimodal distribution across all cuisines, with peaks typically in the range of 7–12

ingredients. Distributions largely overlap, indicating a characteristic compositional scale. Right-skewed tails reflect occasional complex recipes with a larger number of ingredients. **b,** Utensil usage across cuisines: Heat map showing the frequency of utensil usage across different cuisines. A small set of core utensils (e.g., bowl, oven, pan, pot, skillet) dominates, while most tools are infrequent, indicating a shared operational backbone with minor cuisine-specific variation. **c,** Distribution of cooking time: The total preparation and cooking time across recipes follows a broad, right-skewed distribution, with most recipes concentrated between 20 and 60 minutes. Long tails correspond to slow or multi-stage preparations, with consistent patterns across cuisines. **d,** Ingredient category composition: Ingredient usage aggregated into major functional categories (e.g., meat, spices, dish, additive, vegetables, bakery, dairy, fruit, cereals, and condiments) reveals a heterogeneous yet structured distribution. A few dominant categories account for most usage, while many contribute to long-tail diversity. This uneven distribution reflects both the central role of staple ingredients and the combinatorial richness of global cuisines. Together, these observations highlight a balance between diversity and constraint in culinary systems: while cuisines differ in specific ingredients, tools, and traditions, they share common structural properties in recipe size, operational processes, temporal scales, and ingredient composition.

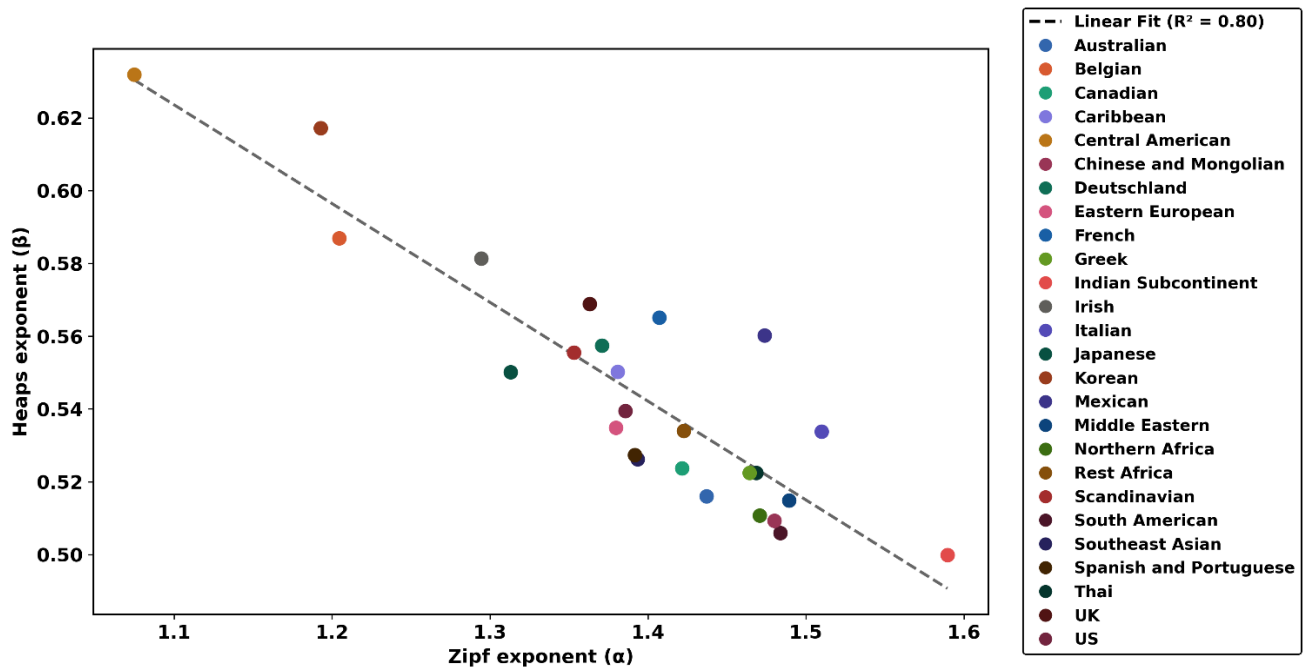

**Extended Data Fig. 3 | Coupling between Zipf and Heaps scaling across global cuisines.** Relationship between the Zipf exponent ($\alpha$) and Heaps exponent ($\beta$) across 26 global cuisines, capturing the interplay between ingredient frequency heterogeneity and the growth of culinary diversity. Each point represents a cuisine, positioned according to its independently estimated scaling exponents. The dashed line indicates a linear fit ($R^2 = 0.80$), revealing a strong inverse relationship between $\alpha$ and $\beta$. Higher values of $\alpha$ correspond to more unequal ingredient usage distributions, where a small number of ingredients dominate recipe composition. In contrast, lower values of $\beta$ indicate slower growth in the number of distinct ingredients as the corpus size increases. The observed negative correlation therefore implies that cuisines characterized by stronger concentration of ingredient usage exhibit reduced rates of vocabulary expansion. Conversely, cuisines with more evenly distributed ingredient usage (lower $\alpha$) tend to explore a broader ingredient space, leading to higher $\beta$. This coupling reflects a fundamental trade-off between reuse and innovation in culinary systems: preferential reuse of common ingredients constrains the introduction of novel ingredients, while greater exploratory diversity reduces dominance effects. Such relationships between Zipf and Heaps exponents have been reported in other complex compositional systems, including natural language and biological sequences, where they emerge from shared generative mechanisms governing element reuse and vocabulary growth. The consistency of this scaling relation across culturally diverse cuisines suggests that culinary systems are governed by universal constraints linking frequency heterogeneity and diversity expansion. This interplay provides a unifying framework connecting the statistical laws of ingredient usage (Zipf's law) and diversity growth (Heaps' law), and supports the existence of common underlying generative processes shaping culinary design.

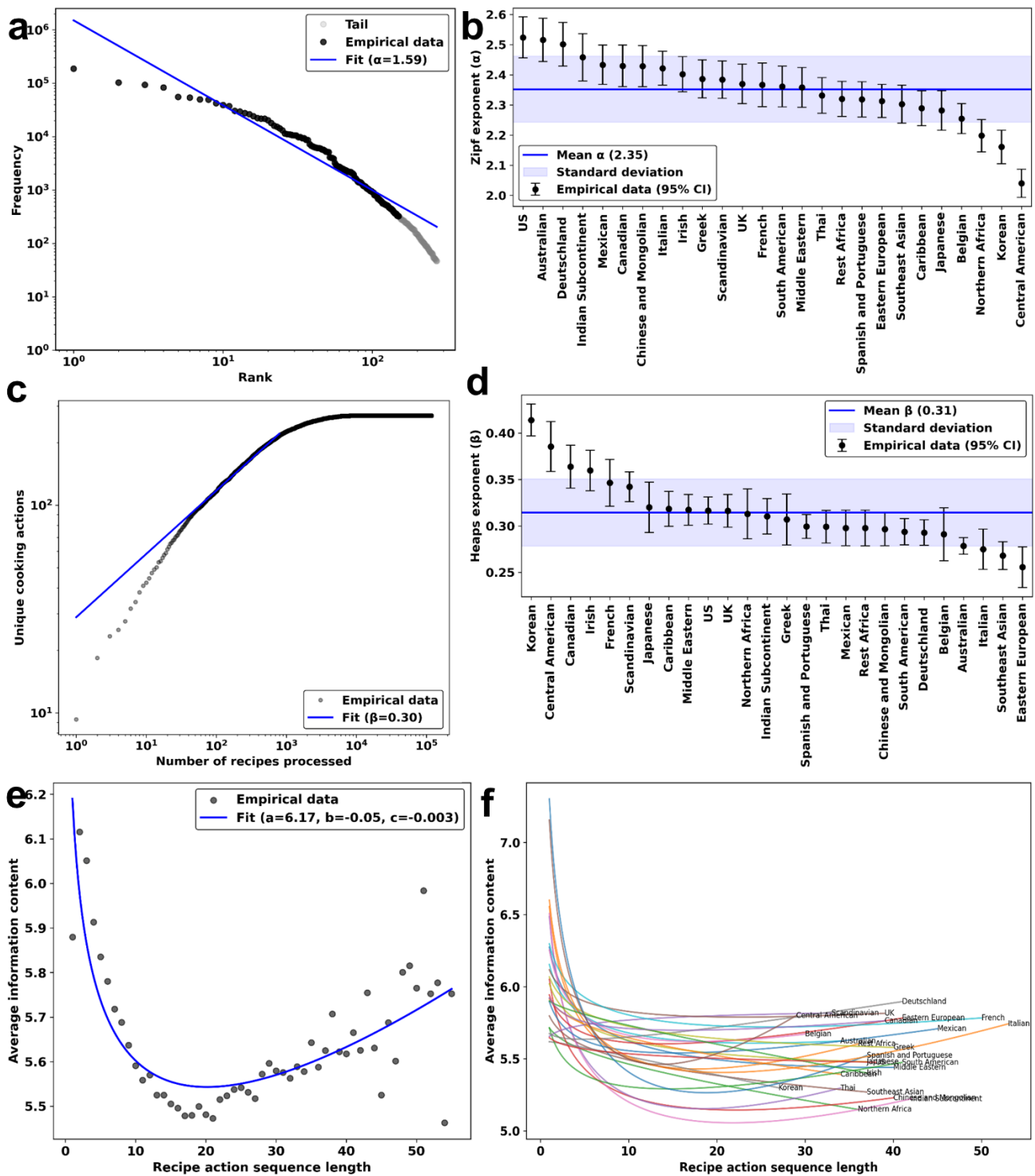

**Extended Data Fig. 4 | Statistical patterns in the organization of cooking techniques across global cuisines.** Cooking actions extracted from recipes exhibit scaling laws analogous to ingredient usage. **a,** Zipf-like rank–frequency distribution: Frequencies follow a heavy-tailed Zipf-like distribution ($f(r) \sim r^{-\alpha}$). The fitted exponent ($\alpha \approx 1.59$) indicates that a small set of common techniques (e.g., mixing, heating, boiling) dominates recipe preparation, while a long tail of rare techniques contributes to diversity. **b,** Cross-cuisine variability of Zipf exponent: Zipf exponents ($\alpha$) vary narrowly across cuisines, suggesting shared preferential reuse mechanisms. **c,** Heaps' law for cooking technique

diversity: The growth of distinct cooking techniques $V(R)$ with increasing number of recipes $R$ follows sublinear scaling ($V(R) \sim R^\beta$, with $\beta \approx 0.30$). This indicates diminishing returns in procedural diversity as more recipes are considered, reflecting constraints on the expansion of the culinary action vocabulary. **d**, Cross-cuisine variability of Heaps exponent: Heaps exponents (β) cluster tightly across cuisines, indicating shared constraints on introducing new techniques. **e,** Menzerath–Altmann relationship in procedural complexity: Average information content varies non-monotonically with sequence length, consistent with Menzerath–Altmann scaling. It decreases initially due to redundancy, then increases with diversity. **f,** Cuisine-specific Menzerath–Altmann fits. Individual cuisines display qualitatively similar non-monotonic profiles, with variations in the position of the minimum reflecting differences in procedural complexity. Collectively, these results demonstrate that cooking techniques—like ingredients—obey statistical laws across cuisines. This parallel structure underscores that both what we cook (ingredients) and how we cook (techniques) are governed by shared principles of organization in complex culinary systems.

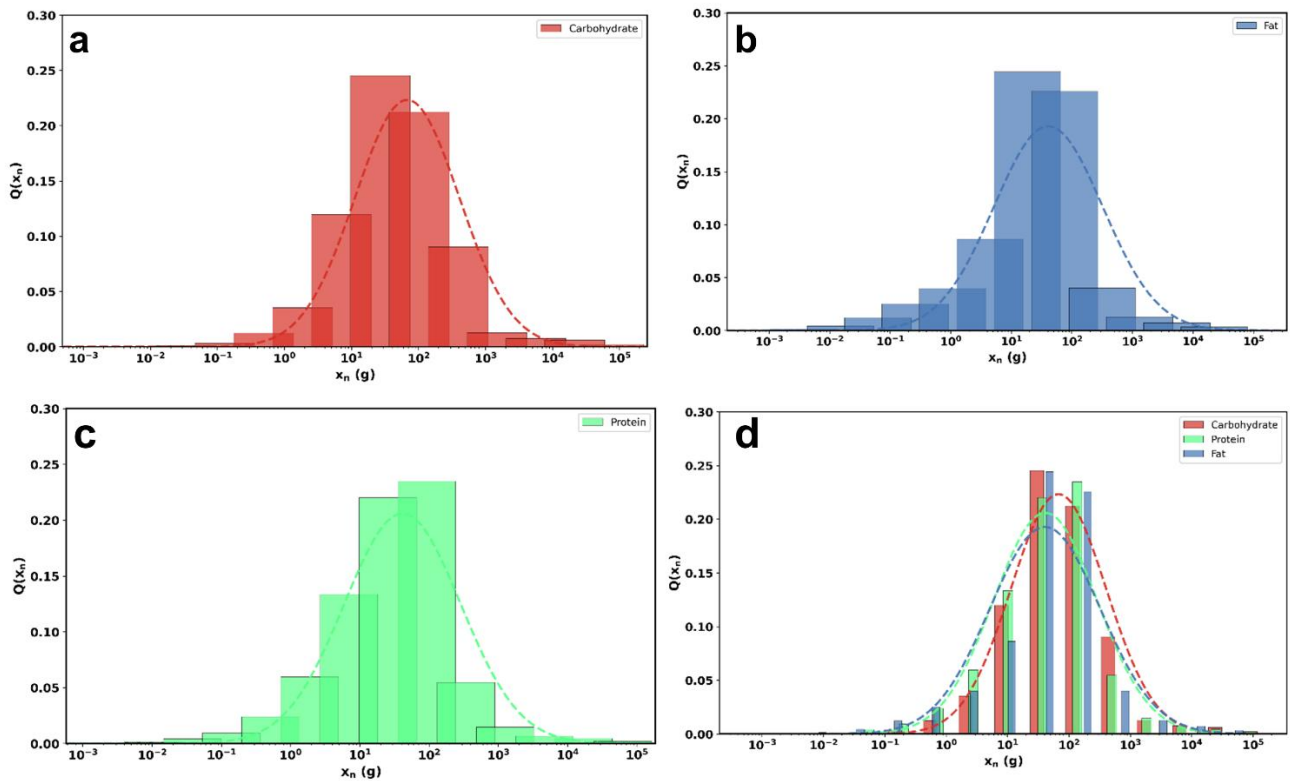

**Extended Data Fig. 5: Log-normal distributions of macronutrient concentrations across global recipes.** Empirical distributions of macronutrient concentrations for carbohydrates, fats, and proteins across the global corpus of recipes spanning 26 cuisines. **a–c,** Macronutrient-specific distributions: Histograms of (a) carbohydrates, (b) total lipids (fats), and (c) proteins are shown on a logarithmic scale of concentration ($x_n$, in grams). In each case, the empirical distributions exhibit right-skewed shapes in linear space that become approximately symmetric and unimodal in log-space, consistent with log-normal behavior. Dashed curves represent fitted log-normal distributions, capturing the central tendency and dispersion of the data across several orders of magnitude. The close agreement between empirical histograms and fitted curves indicates that macronutrient concentrations are well described by multiplicative stochastic processes underlying recipe composition. **d,** Cross-nutrient universality: Overlay of the normalized distributions for carbohydrates, fats, and proteins demonstrates a strong collapse onto a common functional form when viewed in log-space. Despite differences in absolute concentrations and biochemical roles, all three macronutrients follow similar log-normal distributions spanning approximately six orders of magnitude. This collapse highlights a common statistical structure governing nutrient composition across recipes. The emergence of log-normal distributions suggests that macronutrient composition is shaped by multiplicative constraints arising from ingredient combinations and cooking transformations. The consistency of these patterns across diverse cuisines indicates that global culinary systems operate under shared statistical principles, independent of cultural or regional variation.

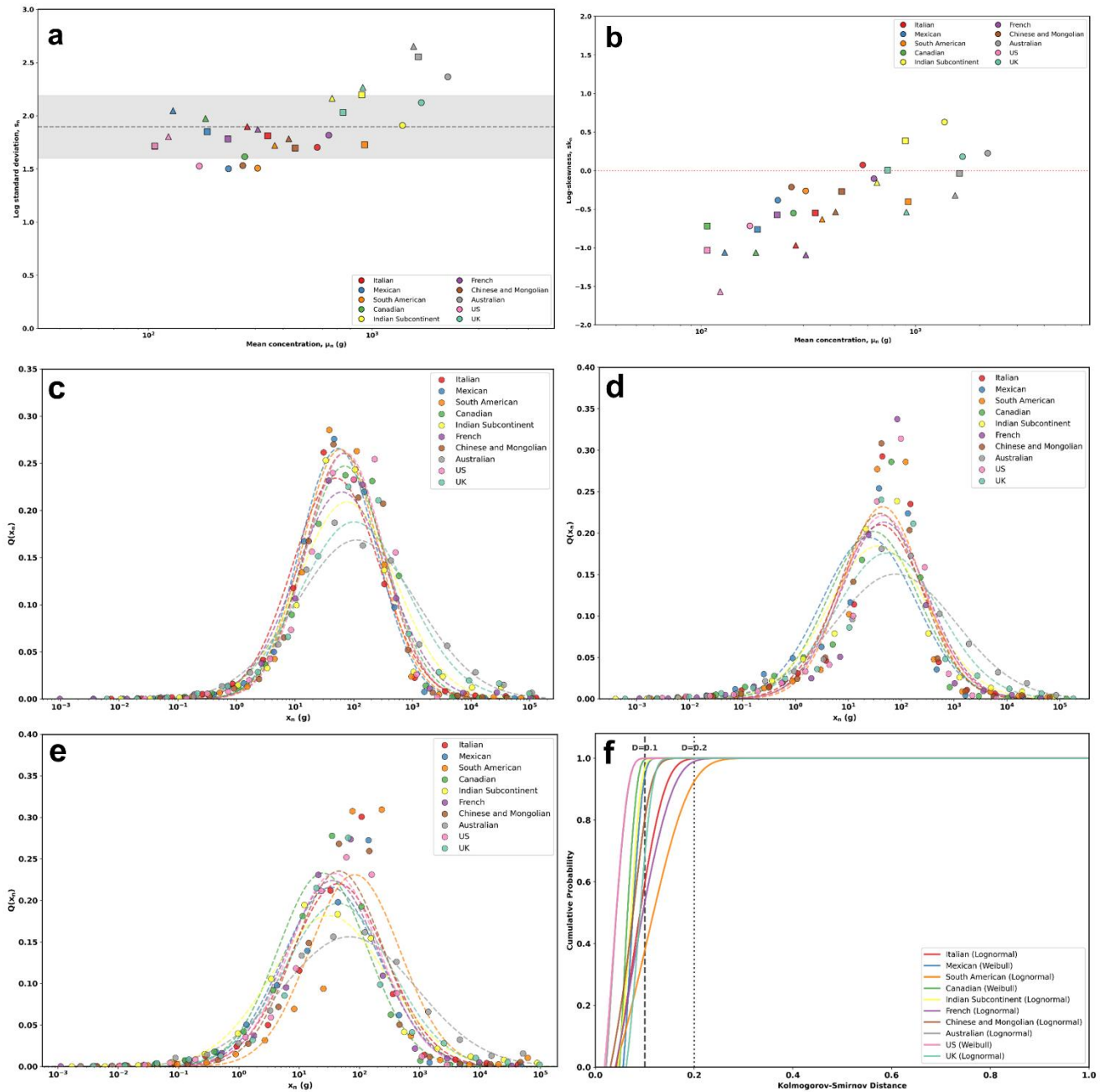

**Extended Data Fig. 6 | Statistical signatures of macronutrient composition across major cuisines.** Analysis of macronutrient concentration distributions (carbohydrates, fats, proteins) across ten major cuisines (Italian, Mexican, South American, Canadian, Indian Subcontinent, French, Chinese and Mongolian, Australian, US, and UK). **a,** Scale-constrained variability: The logarithm of the standard deviation ($s_n$) of macronutrient concentrations is plotted against the mean concentration ($\mu_n$) for each cuisine. Data points for all three macronutrients cluster within a narrow band, indicating limited dispersion across cuisines despite variations in absolute concentration levels. The shaded region denotes the characteristic range of variability, suggesting a shared constraint governing nutrient fluctuations. **b,** Symmetry of distributions: Log-skewness ($s_{k_n}$) of macronutrient concentrations is shown as a function of mean concentration. Values remain close to zero across cuisines, indicating approximate symmetry in log-space. The dashed line marks zero skewness (perfect symmetry), while deviations are small and

bounded, consistent with log-normal distributions. **c–e,** Macronutrient-specific distributions across cuisines: Probability density functions of (c) carbohydrates, (d) fats, and (e) proteins are shown for each cuisine on a logarithmic concentration scale. Empirical distributions collapse onto similar unimodal, symmetric curves in log-space, with dashed lines indicating fitted log-normal distributions. Despite differences in culinary practices and ingredient compositions, all cuisines exhibit comparable distribution shapes spanning several orders of magnitude. **f,** Goodness-of-fit analysis: Kolmogorov–Smirnov statistics for the best model across candidate models (log-normal and alternatives) demonstrate that the log-normal distribution provides superior fits (lower KS distances) for most cuisines. Vertical reference lines indicate representative KS thresholds. Together, these results confirm that the log-normal signature of macronutrient concentrations persists across individual cuisines. The observed consistency—characterized by constrained variability, near-zero skewness, and strong goodness-of-fit—suggests that nutrient composition in culinary systems is governed by multiplicative processes and constraints, independent of cultural or regional variation.

# Universal statistical laws governing culinary design

Ganesh Bagler[1,2,3,6*], Gopal Krishna Tewari[4†], Aditya Raj Yadav[4†], Akshat Singh[5†], Pranay Bansal[5†], Ujjval Dargar[5†], Mansi Goel[1,2,3†], and Madhvi Kumari Sinha[1,2,3†]

[1]Department of Computational Biology
[2]Infosys Center for Artificial Intelligence
[3]Center of Excellence in Healthcare
[4]Department of Mathematics
[5]Department of Computer Science
Indraprastha Institute of Information Technology Delhi (IIIT-Delhi), New Delhi 110020 India.
[6]Foodoscope Technologies Private Limited, New Delhi 110048 India.

*Corresponding Author: bagler@iiitd.ac.in
†Equal contribution

1. **SI Tables**
   - Supplementary Table 1 | Number of recipes across cuisines at the regions-level.
   - Supplementary Table 2 | The geo-cultural mappings of the recipes at the level of continent, region and sub-region (country).
   - Supplementary Table 3 | Performance of the deep-learning based named entity recognition models.
   - Supplementary Table 4 | The list 100 most frequently ingredients across the global recipes corpus.
   - Supplementary Table 5 | The list of cooking techniques extracted from the global recipes corpus using the state-of-the-art NER protocol.
   - Supplementary Table 6 | The Zipf coefficient ($\alpha$) for every cuisine.
   - Supplementary Table 7 | The Heap's coefficient ($\beta$) for every cuisine.
   - Supplementary Table 8 | The Menzerath-Altmann coefficients (a, b, c) for every cuisine.
2. **SI Figures**
   - Supplementary Fig. 1 | Global culinary corpus and recipes as complex compositional systems.
   - Supplementary Fig. 2 | Transformer-based architecture for culinary named entity recognition.
   - Supplementary Fig. 3 | Menzerath–Altmann scaling in culinary design across world cuisines.
   - Supplementary Fig. 4 | Log-normal organization and scale invariance of macronutrient concentrations across major cuisines.
   - Supplementary Fig. 5 | Power-law scaling between mean and variability of macronutrient concentrations.
   - Supplementary Fig. 6 | Kolmogorov-Smirnov statistics for the ten major cuisines.

- Supplementary Fig. 7 | The log-normal distribution of nutrient concentrations in the ten major cuisines.
- Supplementary Fig. 8 | Translational invariance of nutrient concentrations in the ten major cuisines.

# 1. SUPPLEMENTARY INFORMATION TABLES

| Region | Number of Recipes |
|---|---|
| Italian | 16574 |
| Mexican | 14447 |
| South American | 7171 |
| Canadian | 6694 |
| Indian Subcontinent | 6463 |
| French | 6375 |
| Chinese and Mongolian | 5888 |
| Australian | 5819 |
| US | 5025 |
| UK | 4396 |
| Deutschland | 4321 |
| Greek | 4181 |
| Middle Eastern | 3905 |
| Caribbean | 3023 |
| Spanish and Portuguese | 2843 |
| Scandinavian | 2811 |
| Rest Africa | 2739 |
| Thai | 2603 |
| Irish | 2529 |
| Eastern European | 2503 |
| Japanese | 2040 |
| Southeast Asian | 1940 |
| Northern Africa | 1610 |
| Belgian | 1060 |
| Korean | 668 |
| Central American | 455 |
| **Total** | **118083** |

**Supplementary Table 1 | Number of recipes across cuisines at the regions-level.** The corpus is representative of diverse global culinary practices.

| Continent | Region | Sub-region (Country) |
|---|---|---|
| African | Middle Eastern | Egyptian |
| African | Northern Africa | Libyan |
| African | Northern Africa | Moroccan |
| African | Rest Africa | Angolan |
| African | Rest Africa | Ethiopian |
| African | Rest Africa | Namibian |
| African | Rest Africa | Nigerian |
| African | Rest Africa | Somalian |
| African | Rest Africa | Sudanese |
| Asian | Chinese and Mongolian | Chinese |
| Asian | Chinese and Mongolian | Mongolian |
| Asian | Indian Subcontinent | Bangladeshi |
| Asian | Indian Subcontinent | Indian |
| Asian | Indian Subcontinent | Nepalese |
| Asian | Indian Subcontinent | Pakistani |
| Asian | Japanese | Japanese |
| Asian | Korean | Korean |
| Asian | Middle Eastern | Iraqi |
| Asian | Middle Eastern | Israeli |
| Asian | Middle Eastern | Laotian |
| Asian | Middle Eastern | Lebanese |
| Asian | Middle Eastern | Palestinian |
| Asian | Middle Eastern | Rest Middle Eastern |
| Asian | Middle Eastern | Saudi Arabian |
| Asian | Middle Eastern | Turkish |
| Asian | Southeast Asian | Cambodian |
| Asian | Southeast Asian | Filipino |
| Asian | Southeast Asian | Indonesian |
| Asian | Southeast Asian | Malaysian |
| Asian | Southeast Asian | Vietnamese |
| Asian | Thai | Thai |
| Australasian | Australian | Australian |
| Australasian | Australian | New Zealander |
| European | Belgian | Belgian |
| European | Belgian | Dutch |
| European | Deutschland | Austrian |
| European | Deutschland | German |
| European | Deutschland | Swiss |
| European | Eastern European | Czech |
| European | Eastern European | Hungarian |
| European | Eastern European | Polish |
| European | Eastern European | Rest Eastern European |
| European | Eastern European | Russian |
| European | French | French |
| European | Greek | Greek |
| European | Irish | Irish |
| European | Italian | Italian |
| European | Scandinavian | Danish |
| European | Scandinavian | Finnish |
| European | Scandinavian | Icelandic |
| European | Scandinavian | Norwegian |
| European | Scandinavian | Swedish |
| European | Spanish and Portuguese | Portuguese |
| European | Spanish and Portuguese | Spanish |
| European | UK | English |
| European | UK | Scottish |
| European | UK | UK |
| European | UK | Welsh |
| Latin American | Caribbean | Cuban |
| Latin American | Caribbean | Jamaican |
| Latin American | Caribbean | Puerto Rican |
| Latin American | Caribbean | Rest Caribbean |
| Latin American | Central American | Costa Rican |
| Latin American | Central American | Guatemalan |
| Latin American | Central American | Honduran |
| Latin American | Mexican | Mexican |
| Latin American | South American | Argentine |
| Latin American | South American | Brazilian |
| Latin American | South American | Chilean |
| Latin American | South American | Colombian |
| Latin American | South American | Ecuadorean |
| Latin American | South American | Peruvian |
| Latin American | South American | Venezuelan |
| North American | Canadian | Canadian |
| North American | US | US |

**Supplementary Table 2 | The geo-cultural mappings of the recipes at the level of continent, region and sub-region (country).** Each cuisine is represented as continent, region, and country (sub-region). The analysis presented here at the level of region.

| Modelling Technique | Manually Annotated | | | Augmented | | | Machine-Annotated | | |
| --- | --- | --- | --- | --- | --- | --- | --- | --- | --- |
| | F1 (%) | P (%) | R (%) | F1 (%) | P (%) | R (%) | F1 (%) | P (%) | R (%) |
| **spaCy-transformer** | **95.90** | **95.89** | **95.91** | **96.04** | **96.05** | **96.04** | **95.71** | **95.73** | **95.69** |
| spaCy-CPU optimized | 94.46 | 94.52 | 94.41 | 94.91 | 94.92 | 94.90 | 91.30 | 91.36 | 91.24 |
| Stanford NER | 95.52 | 95.64 | 95.39 | 95.16 | 94.37 | 95.96 | 89.9 | 91.31 | 88.53 |
| DistilBERT | 93.80 | 95.20 | 93.60 | 93.50 | 93.50 | 94.60 | 90.20 | 92.20 | 89.70 |
| BERT | 94.00 | 94.70 | 94.10 | 93.60 | 93.70 | 94.10 | 90.30 | 91.50 | 90.20 |
| DistilRoBERTa | 93.80 | 94.80 | 93.90 | 94.60 | 94.10 | 95.90 | 90.60 | 91.60 | 90.60 |
| RoBERTa | 92.40 | 92.90 | 92.60 | 94.00 | 94.50 | 94.10 | 90.40 | 91.60 | 90.20 |
| flair | 95.01 | 96.11 | 96.05 | 94.45 | 95.87 | 96.14 | 89.85 | 88.71 | 89.22 |

**Supplementary Table 3 | Performance of the deep-learning based named entity recognition models.** Overall, eight NER models were implemented and their performance was compared using F1 score, precision, and recall. The best results were obtained with the spaCy-transformer which was used for extracting named entities from the recipe text. *Adapted from Goel et al., "Deep Learning Based Named Entity Recognition Models for Recipes", Proceedings of LREC-COLING (2024), 4542–4554.*

| | | | |
|---|---|---|---|
| salt | parmesan cheese | nutmeg | egg yolk |
| onion | soy sauce | vanilla extract | vanilla |
| butter | beef | cayenne pepper | sesame oil |
| water | oregano | red pepper flake | shallot |
| garlic clove | potato | honey | green pepper |
| olive oil | green onion | white wine | chicken stock |
| sugar | brown sugar | coriander | scallion |
| egg | oil | red bell pepper | zucchini |
| tomato | basil | tomato paste | salsa |
| black pepper | chicken broth | salt black pepper | spinach |
| garlic | lemon | chicken | canola oil |
| milk | lime juice | mozzarella cheese | green chilies |
| pepper | extra virgin olive oil | tomato sauce | shrimp |
| salt pepper | chicken breast | almond | sea salt |
| flour | white sugar | clove | breadcrumb |
| cinnamon | mushroom | kosher salt | orange juice |
| parsley | paprika | bacon | lime |
| lemon juice | bay leaf | worcestershire sauce | raisin |
| purpose flour | chili powder | red pepper | red wine |
| vegetable oil | garlic powder | green bell pepper | cucumber |
| cumin | thyme | turmeric | jalapeno pepper |
| carrot | cornstarch | cream cheese | beef broth |
| cream | celery | rice | coconut milk |
| ginger | red onion | heavy cream | egg white |
| cilantro | cheddar cheese | mayonnaise | curry powder |

**Supplementary Table 4 | The list of 100 most frequently ingredients across the global recipes corpus.** The NER model identified 19,019 unique ingredients, including variations in different forms of primary ingredients.

| | | | | | | | |
|---|---|---|---|---|---|---|---|
| add | shape | wok | defrost | fold | scatter | sieve | tilt |
| heat | wrap | split | splutter | dry | reheat | hold | lard |
| cook | soak | rest | scramble | slice | flip | swirl | deflate |
| stir | fill | baste | thicken | coat | dredge | poke | sort |
| place | roast | bubble | scald | cream | crush | charred | strip |
| mix | broil | burn | twist | chop | clean | floured | submerge |
| cover | stir-fry | skin | barbecue | taste | check | sizzle | dollop |
| remove | dust | pop | unmold | peel | stuff | wet | rotate |
| serve | drop | squash | poach | garnish | grate | dump | splash |
| boil | dip | move | presentation | brush | whip | mound | pan-fry |
| simmer | marinate | soften | dress | press | bubbling | frost | curdle |
| bake | ladle | blanch | smear | drizzle | punch | test | freezing |
| stirring | scoop | raise | crockpot | roll | warm | thaw | stop |
| sprinkle | dice | deglaze | overmix | grease | skim | distribute | char |
| cool | mash | pierce | smash | chill | smoking | wait | slash |
| preheat | uncover | stream | devein | saute | thread | glaze | crimp |
| cut | dissolve | absorbed | powdered | prepare | wipe | sit | muddle |
| combine | separate | assemble | moist | knead | decorate | scrub | sterilize |
| drain | dressing | stack | evaporate | stand | marinade | carve | blot |
| pour | uncovered | moisten | flake | process | zest | drip | foam |
| set | pat | ice | roux | divide | handle | overcook | flavoring |
| season | shake | replace | slit | toss | rub | stew | touch |
| beat | turn | pack | dripping | strain | flatten | start | snap |
| spread | take | foil | repeat | puree | pull | style | dash |
| transfer | break | tear | parboil | rinse | push | select | immerse |
| whisk | grind | square | sprout | spray | reserve | griddle | massage |
| smooth | toast | caramelized | unfold | seasoning | mince | unroll | settle |
| top | evaporated | smoke | rising | wash | prick | insert | sweeten |
| blend | steam | throw | stretch | arrange | open | sweat | wilt |
| refrigerate | squeeze | yield | steep | store | pre-heat | tuck | slather |
| put | crumble | save | braise | sift | butter | meld | mould |
| fry | scrape | tie | condensed | seal | sear | unwrap | blitz |
| melt | shred | pick | whirl | rise | measure | scorching | |
| reduce | trim | invert | minced | note | pressure | caramelize | |

**Supplementary Table 5 | The list of cooking techniques extracted from the global recipes corpus using the state-of-the-art NER protocol.** It techniques span across physical, chemical, and biological transformations brought about in the ingredients through act of cooking.

| Cuisine | α | Lower_CI_95 | Upper_CI_95 | $R^2$ |
|---|---|---|---|---|
| Indian Subcontinent | 1.590 | 1.579 | 1.601 | 0.969 |
| Italian | 1.510 | 1.502 | 1.518 | 0.963 |
| Middle Eastern | 1.489 | 1.478 | 1.501 | 0.970 |
| South American | 1.484 | 1.475 | 1.493 | 0.969 |
| Chinese and Mongolian | 1.480 | 1.471 | 1.489 | 0.971 |
| Mexican | 1.474 | 1.466 | 1.482 | 0.961 |
| Northern Africa | 1.471 | 1.457 | 1.484 | 0.974 |
| Thai | 1.469 | 1.456 | 1.481 | 0.966 |
| Greek | 1.464 | 1.453 | 1.476 | 0.968 |
| Australian | 1.437 | 1.428 | 1.446 | 0.970 |
| Rest Africa | 1.423 | 1.411 | 1.435 | 0.968 |
| Canadian | 1.422 | 1.414 | 1.430 | 0.973 |
| French | 1.407 | 1.399 | 1.416 | 0.970 |
| Southeast Asian | 1.394 | 1.381 | 1.406 | 0.968 |
| Spanish and Portuguese | 1.392 | 1.381 | 1.403 | 0.969 |
| US | 1.386 | 1.377 | 1.394 | 0.973 |
| Caribbean | 1.381 | 1.371 | 1.391 | 0.969 |
| Eastern European | 1.380 | 1.368 | 1.392 | 0.968 |
| Deutschland | 1.371 | 1.362 | 1.380 | 0.971 |
| UK | 1.363 | 1.354 | 1.373 | 0.967 |
| Scandinavian | 1.353 | 1.342 | 1.365 | 0.970 |
| Japanese | 1.313 | 1.302 | 1.324 | 0.968 |
| Irish | 1.294 | 1.284 | 1.305 | 0.966 |
| Belgian | 1.205 | 1.192 | 1.218 | 0.967 |
| Korean | 1.193 | 1.177 | 1.209 | 0.962 |
| Central American | 1.075 | 1.059 | 1.090 | 0.956 |

**Supplementary Table 6 | The Zipf coefficient ($\alpha$) for every cuisine**. The 95% lower and upper confidence intervals and goodness of fit ($R^2$) are also reported for each cuisine.

| Cuisine | β | Lower_CI_95 | Upper_CI_95 | $R^2$ | Num_Recipes |
|---|---|---|---|---|---|
| Central American | 0.632 | 0.617 | 0.647 | 0.994 | 455 |
| Korean | 0.617 | 0.606 | 0.629 | 0.994 | 668 |
| Belgian | 0.587 | 0.577 | 0.597 | 0.999 | 1060 |
| Irish | 0.581 | 0.577 | 0.586 | 0.999 | 2529 |
| UK | 0.569 | 0.565 | 0.573 | 0.999 | 4396 |
| French | 0.565 | 0.561 | 0.570 | 0.998 | 6375 |
| Mexican | 0.560 | 0.559 | 0.562 | 1.000 | 14447 |
| Deutschland | 0.557 | 0.552 | 0.562 | 0.998 | 4321 |
| Scandinavian | 0.556 | 0.547 | 0.564 | 0.997 | 2811 |
| Caribbean | 0.550 | 0.545 | 0.556 | 0.999 | 3023 |
| Japanese | 0.550 | 0.541 | 0.559 | 0.997 | 2040 |
| US | 0.540 | 0.533 | 0.546 | 0.996 | 5025 |
| Eastern European | 0.535 | 0.528 | 0.542 | 0.998 | 2503 |
| Rest Africa | 0.534 | 0.528 | 0.540 | 0.999 | 2739 |
| Italian | 0.534 | 0.533 | 0.535 | 1.000 | 16574 |
| Spanish and Portuguese | 0.527 | 0.523 | 0.532 | 0.999 | 2843 |
| Southeast Asian | 0.526 | 0.521 | 0.531 | 0.999 | 1940 |
| Canadian | 0.524 | 0.519 | 0.529 | 0.997 | 6694 |
| Greek | 0.523 | 0.520 | 0.525 | 0.999 | 4181 |
| Thai | 0.523 | 0.519 | 0.526 | 0.999 | 2603 |
| Australian | 0.516 | 0.511 | 0.521 | 0.998 | 5819 |
| Middle Eastern | 0.515 | 0.512 | 0.518 | 0.999 | 3905 |
| Northern Africa | 0.511 | 0.502 | 0.520 | 0.998 | 1610 |
| Chinese and Mongolian | 0.509 | 0.506 | 0.513 | 0.999 | 5888 |
| South American | 0.506 | 0.503 | 0.509 | 0.999 | 7171 |
| Indian Subcontinent | 0.500 | 0.498 | 0.502 | 0.999 | 6463 |

**Supplementary Table 7 | The Heap's coefficient ($\beta$) for every cuisine.** The 95% lower and upper confidence intervals, goodness of fit ($R^2$), and number of recipes are also reported for each cuisine.

| Cuisine | a | b | c |
|---|---|---|---|
| Middle Eastern | 9.227 | -0.124 | -0.012 |
| Rest Africa | 11.122 | -0.207 | -0.015 |
| Northern Africa | 10.059 | -0.184 | -0.012 |
| Chinese and Mongolian | 10.536 | -0.196 | -0.017 |
| Thai | 11.731 | -0.271 | -0.022 |
| Southeast Asian | 8.059 | -0.008 | -0.001 |
| Indian Subcontinent | 10.244 | -0.149 | -0.009 |
| Korean | 10.798 | -0.356 | -0.039 |
| Japanese | 8.532 | -0.089 | -0.013 |
| Australian | 12.182 | -0.248 | -0.022 |
| Mexican | 10.787 | -0.181 | -0.014 |
| Caribbean | 12.093 | -0.271 | -0.022 |
| South American | 12.484 | -0.269 | -0.020 |
| Eastern European | 11.225 | -0.295 | -0.029 |
| Scandinavian | 11.918 | -0.393 | -0.042 |
| UK | 11.958 | -0.308 | -0.031 |
| Belgian | 11.892 | -0.396 | -0.044 |
| Deutschland | 12.379 | -0.348 | -0.034 |
| Greek | 12.487 | -0.339 | -0.028 |
| French | 12.434 | -0.314 | -0.029 |
| Spanish and Portuguese | 11.337 | -0.253 | -0.022 |
| Italian | 12.092 | -0.249 | -0.017 |
| Irish | 11.608 | -0.253 | -0.023 |
| US | 11.872 | -0.252 | -0.021 |
| Canadian | 12.248 | -0.272 | -0.024 |
| Central American | 10.173 | -0.156 | -0.013 |

**Supplementary Table 8 | The Menzerath-Altmann coefficients ($a$, $b$, $c$) for every cuisine.**

## 2. SUPPLEMENTARY INFORMATION FIGURES

**a**

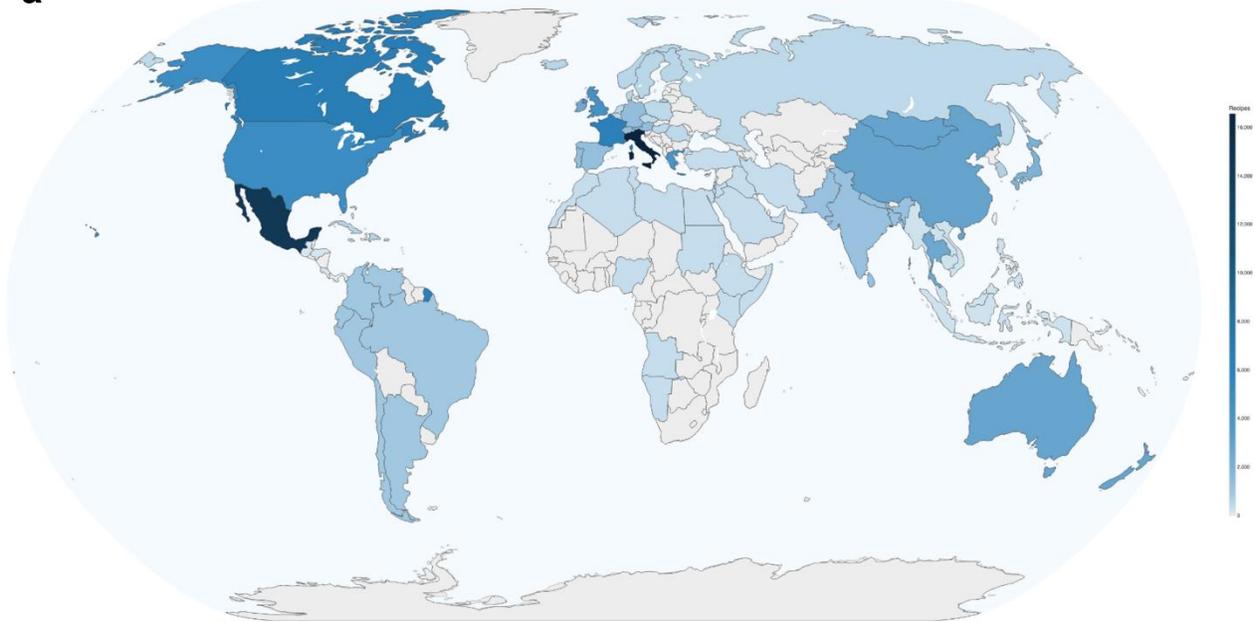

**b**

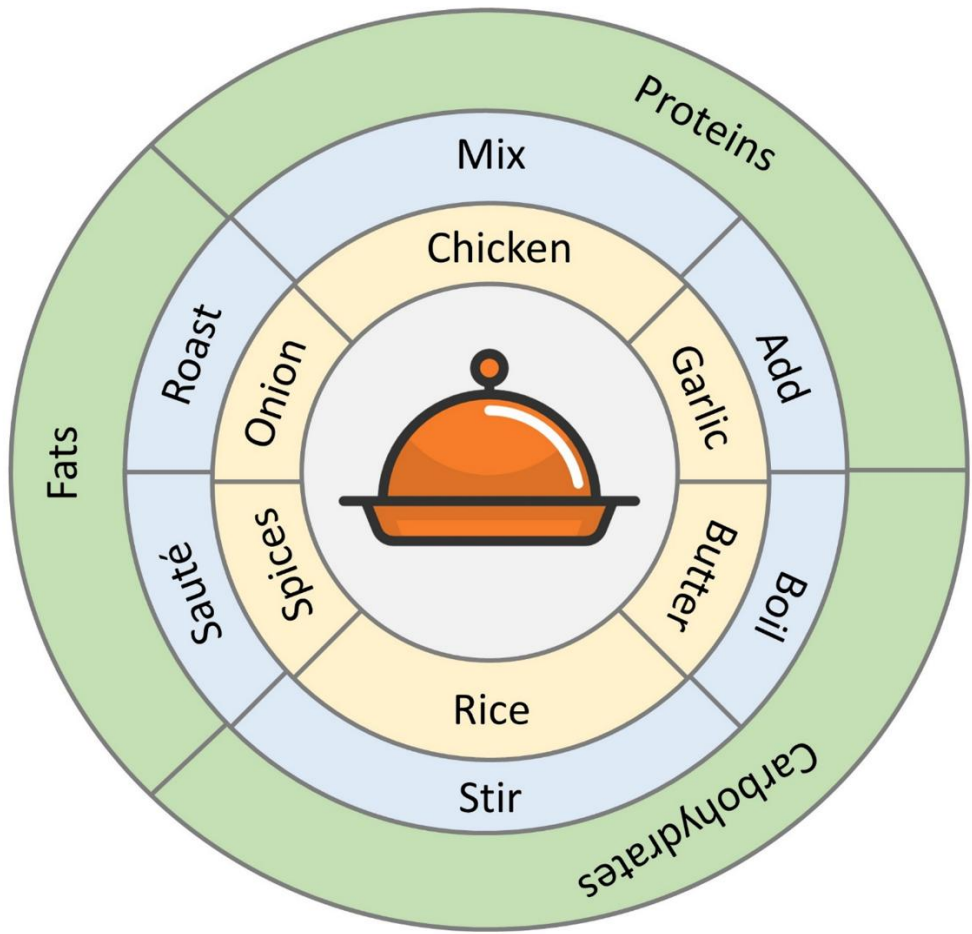

**Supplementary Fig. 1 | Global culinary corpus and recipes as complex compositional systems. a,** Global distribution of the culinary dataset: The recipe corpus comprises 118,083 recipes spanning 26 regional cuisines across 75 countries, representing a broad spectrum of geographical, cultural, and climatic diversity. The dataset includes cuisines from major world regions: Middle Eastern, Northern Africa, Rest Africa, Chinese and Mongolian, Indian Subcontinent, Japanese, Korean, Middle Eastern, Southeast Asian, Thai, Australian, Belgian, Deutschland, Eastern European, French, Greek, Irish, Italian, Scandinavian, Spanish and Portuguese, UK, Caribbean, Central American, Mexican, South American, Canadian, and US. The global coverage ensures that the analysis captures diverse culinary traditions and ingredient ecosystems. **b,** Recipe as a structured symbolic system: A recipe is conceptualized as a compositional system arising from the structured combination of ingredients through sequences of culinary transformations. Ingredients (e.g., rice, chicken, spices) interact with cooking processes (e.g., chopping, mixing, roasting, boiling), resulting in a final dish characterized by its nutritional and sensory attributes. By extracting culinary named entities—including ingredients, quantities, and cooking actions—from this large corpus, the framework enables systematic investigation of statistical regularities underlying culinary design across cultures.

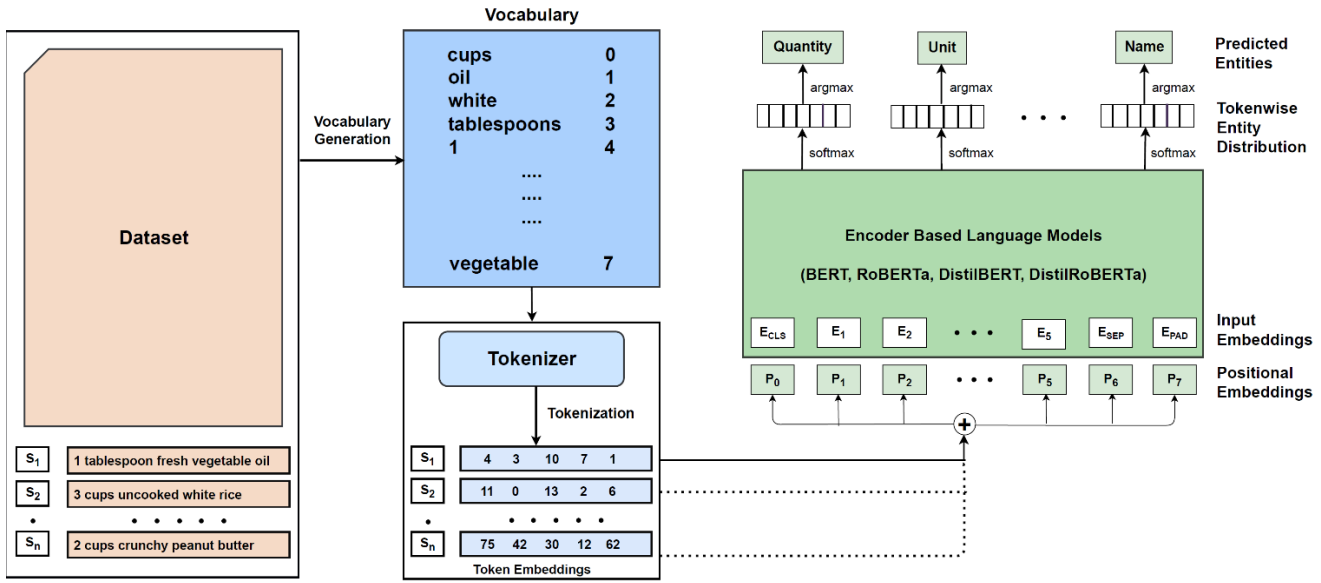

**Supplementary Fig. 2 | Transformer-based architecture for culinary named entity recognition.** Schematic representation of the supervised deep learning pipeline employed for fine-tuning state-of-the-art transformer-based models for named entity recognition in recipes. The framework consists of three major stages. (1) Vocabulary construction and dataset preparation. Raw recipe datasets—comprising ingredient phrases and instruction text—are processed to generate task-specific vocabularies. Each unique token is assigned a numerical index, forming the basis for downstream tokenization. Example input sequences (e.g., "1 tablespoon fresh vegetable oil") illustrate the diversity and compositional nature of culinary text. (2) Tokenization and embedding generation. Input sentences are converted into sequences of token indices using the constructed vocabularies. These token sequences are then mapped to dense vector representations (token embeddings), preserving semantic and syntactic information. Positional embeddings are incorporated to encode token order, enabling the model to capture sequential dependencies within recipe text. (3) Transformer-based encoding and entity prediction. The embedded sequences are processed using encoder-only transformer architectures, including BERT, RoBERTa, DistilBERT, and DistilRoBERTa. Contextualized token representations are obtained through self-attention mechanisms. For each token, the model outputs a probability distribution over entity labels (e.g., ingredient name, quantity, unit), computed via a softmax layer, with final predictions obtained through argmax decoding. Among the evaluated models, the spaCy-transformer pipeline achieved the best performance, yielding F1 scores of 95.9%, 96.04%, and 95.71% on the manually annotated, augmented, and machine-annotated datasets, respectively. The high accuracy underscores the effectiveness of transformer-based architectures in capturing the structured semantics of culinary language. *Adapted from Goel et al., "Deep Learning Based Named Entity Recognition Models for Recipes", Proceedings of LREC-COLING (2024), 4542–4554.*

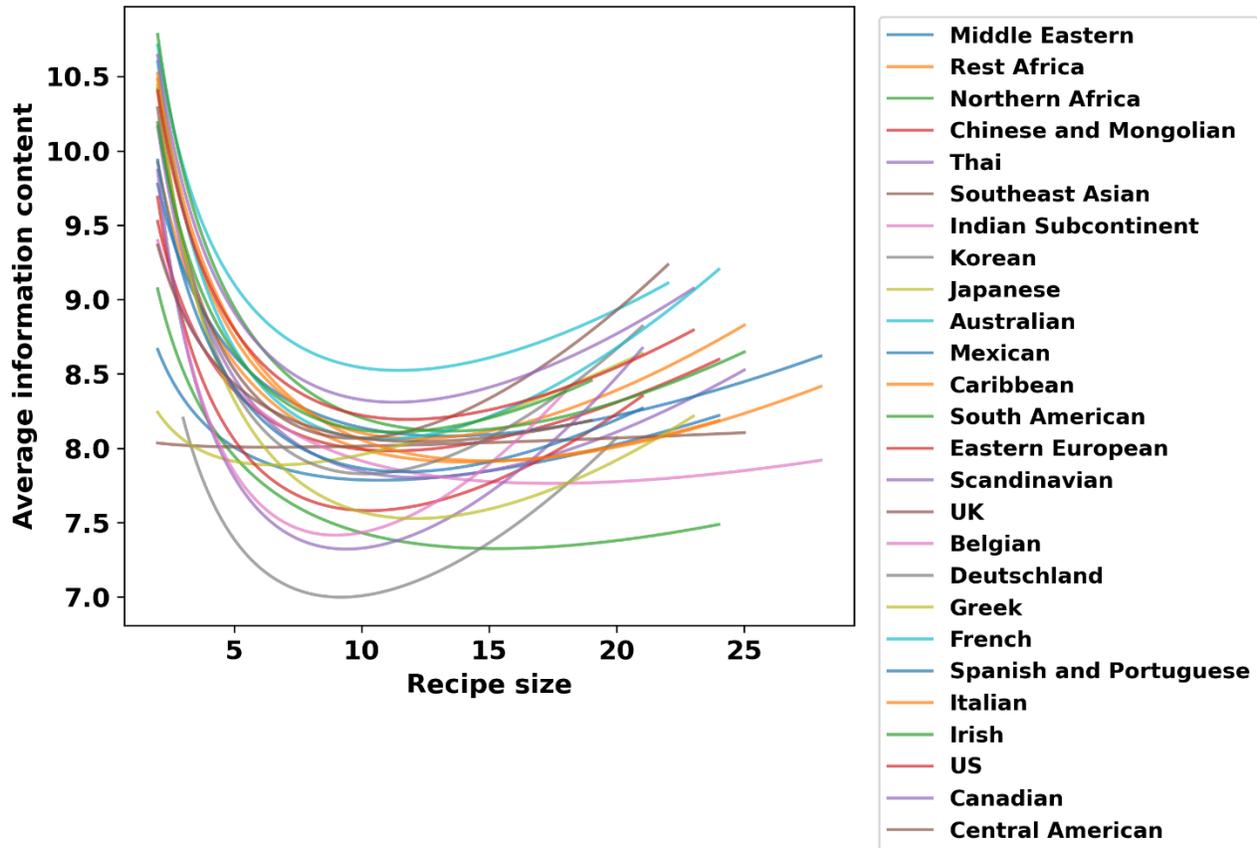

**Supplementary Fig. 3 | Menzerath–Altmann scaling in culinary design across world cuisines.** The relationship between recipe size (number of ingredients, $L$) and the corresponding average information content per ingredient ($y(L)$) is shown for 26 regional cuisines. Each curve represents the fitted Menzerath–Altmann functional form, capturing how compositional complexity varies with increasing recipe size. Across all cuisines, recipe complexity exhibits a characteristic non-monotonic dependence on size, consistent with Menzerath–Altmann-type scaling observed in other complex compositional systems. At small sizes, the addition of ingredients leads to a reduction in average information content, reflecting increased redundancy and preferential reuse of common ingredients. This regime suggests an efficiency-driven organization, where frequently used ingredients dominate early growth. Beyond an intermediate regime—typically corresponding to medium-sized recipes—the trend reverses, and average information content begins to increase with size. This rise reflects the growing influence of combinatorial interactions and ingredient diversity, where less frequent ingredients and novel combinations contribute disproportionately to complexity. Despite substantial cultural and regional differences, the qualitative form of the scaling relation is conserved across cuisines, indicating a universal organizational principle governing culinary composition. The observed trade-off between redundancy and diversity highlights an inherent balance between efficiency and expressivity in recipe design, analogous to patterns reported in language, music, and other structured systems.

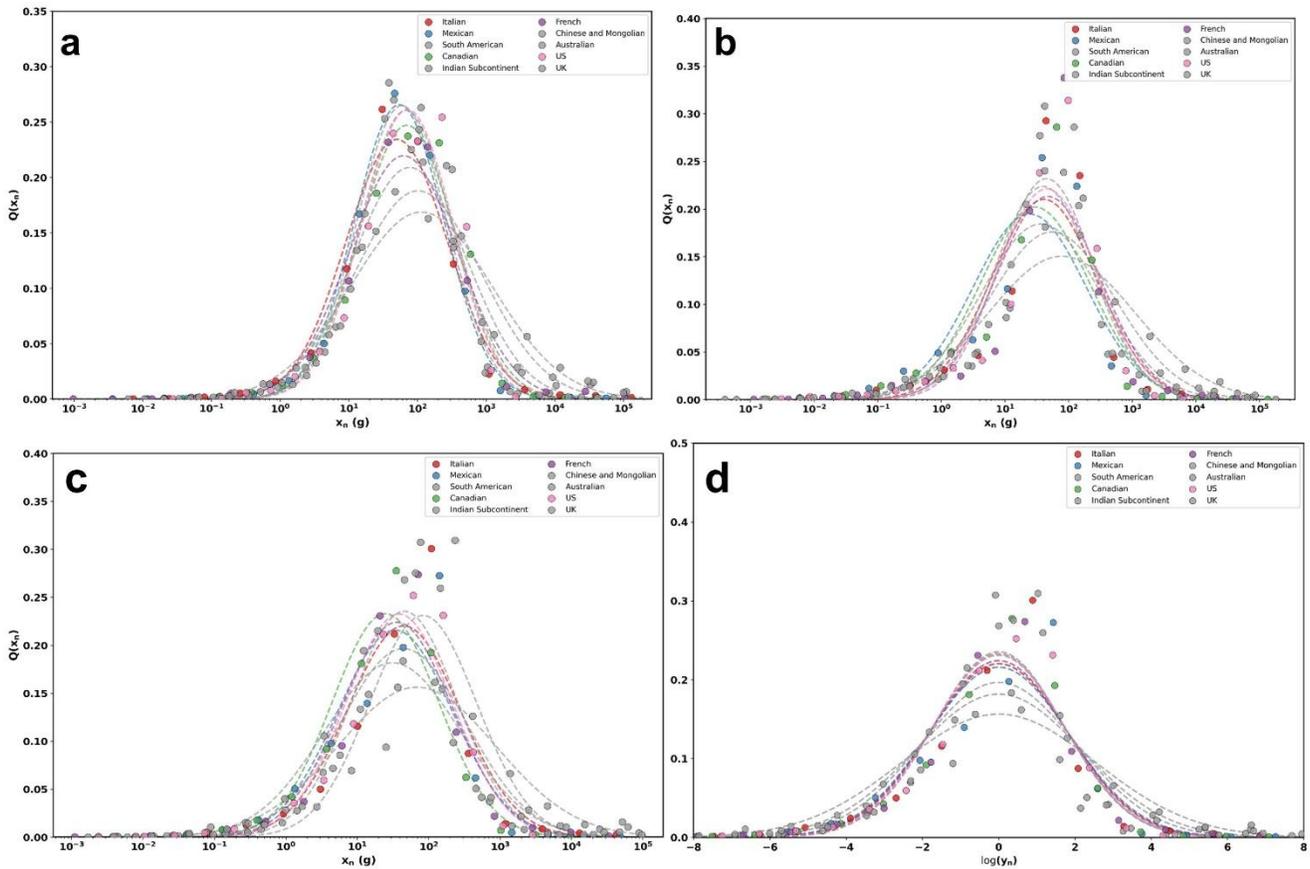

**Supplementary Fig. 4 | Log-normal organization and scale invariance of macronutrient concentrations across major cuisines.** Distributions of macronutrient concentrations—carbohydrates, fats, and proteins—are shown for recipes drawn from ten of the largest regional cuisines (Argentine, Australian, Canadian, Chinese and Mongolian, French, Greek, Indian Subcontinent, Italian, Mexican, and US). **a–c,** Macronutrient-specific distributions: Probability density functions of (a) carbohydrates, (b) fats, and (c) proteins are plotted on a logarithmic concentration scale ($x_n$, in grams). Each cuisine exhibits a unimodal and approximately symmetric distribution in log-space, well described by log-normal fits (dashed curves). Despite substantial diversity in ingredients, culinary practices, and regional contexts, the distributions collapse onto similar functional forms, spanning multiple orders of magnitude in concentration. This consistency indicates that macronutrient composition in recipes is governed by multiplicative processes and shared constraints across cuisines. **d,** Translational invariance in log-space: When expressed in terms of normalized log-concentrations ($\log(y_n)$), the distributions align across cuisines, differing primarily by shifts in location (mean) rather than shape. This translational invariance implies that the underlying generative mechanisms preserve the functional form of the distribution while allowing cuisine-specific scaling of nutrient levels. The collapse of distributions in this representation highlights statistical structure underlying culinary composition. Together, these results demonstrate that nutrient concentrations in recipes exhibit log-normal scaling and invariance properties, reinforcing the existence of organizational principles in culinary systems that transcend cultural and geographical boundaries.

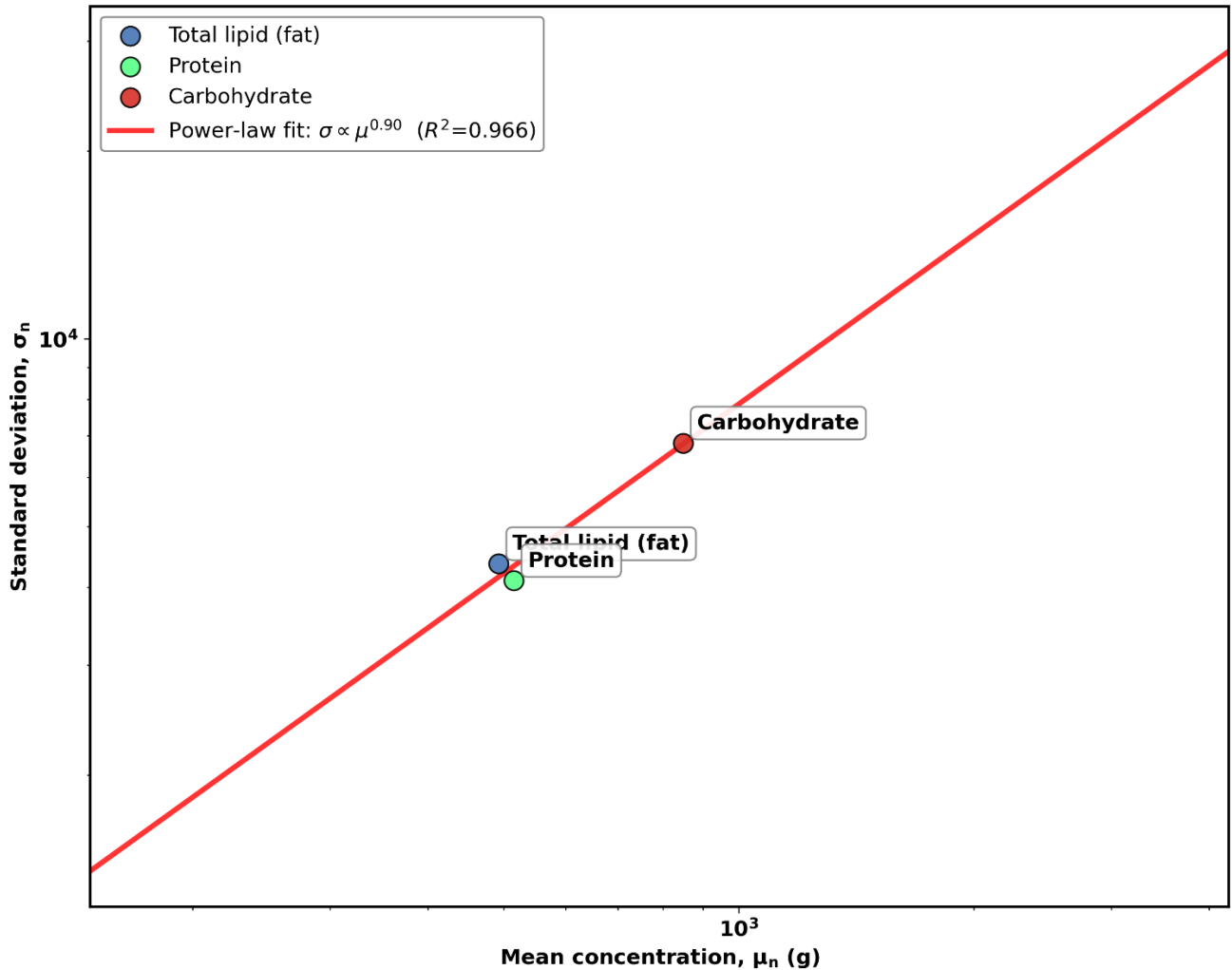

**Supplementary Fig. 5 | Power-law scaling between mean and variability of macronutrient concentrations.** The relationship between the mean concentration ($\mu_n$) and the corresponding standard deviation ($\sigma_n$) of macronutrient distributions—carbohydrates, proteins, and total lipids (fats)—is shown for the global recipe corpus. Each point represents a macronutrient aggregated across all cuisines, plotted on logarithmic axes. A strong scaling relation is observed, with the standard deviation increasing as a power-law function of the mean, $\sigma_n \propto \mu_n^{0.9009}$ (red line; $R^2 = 0.9656$). This near-linear trend in log–log space indicates that fluctuations in nutrient concentrations scale proportionally with their mean values. Such scaling behavior is a hallmark of log-normal distributions, where multiplicative processes govern variability. The observed exponent close to unity suggests that relative variability remains approximately constant across nutrients, implying scale-invariant organization of nutrient concentrations in recipes. This result provides an independent validation of the log-normal framework observed in the distributional analyses and reinforces the hypothesis that multiplicative mechanisms underlie nutrient composition in culinary systems.

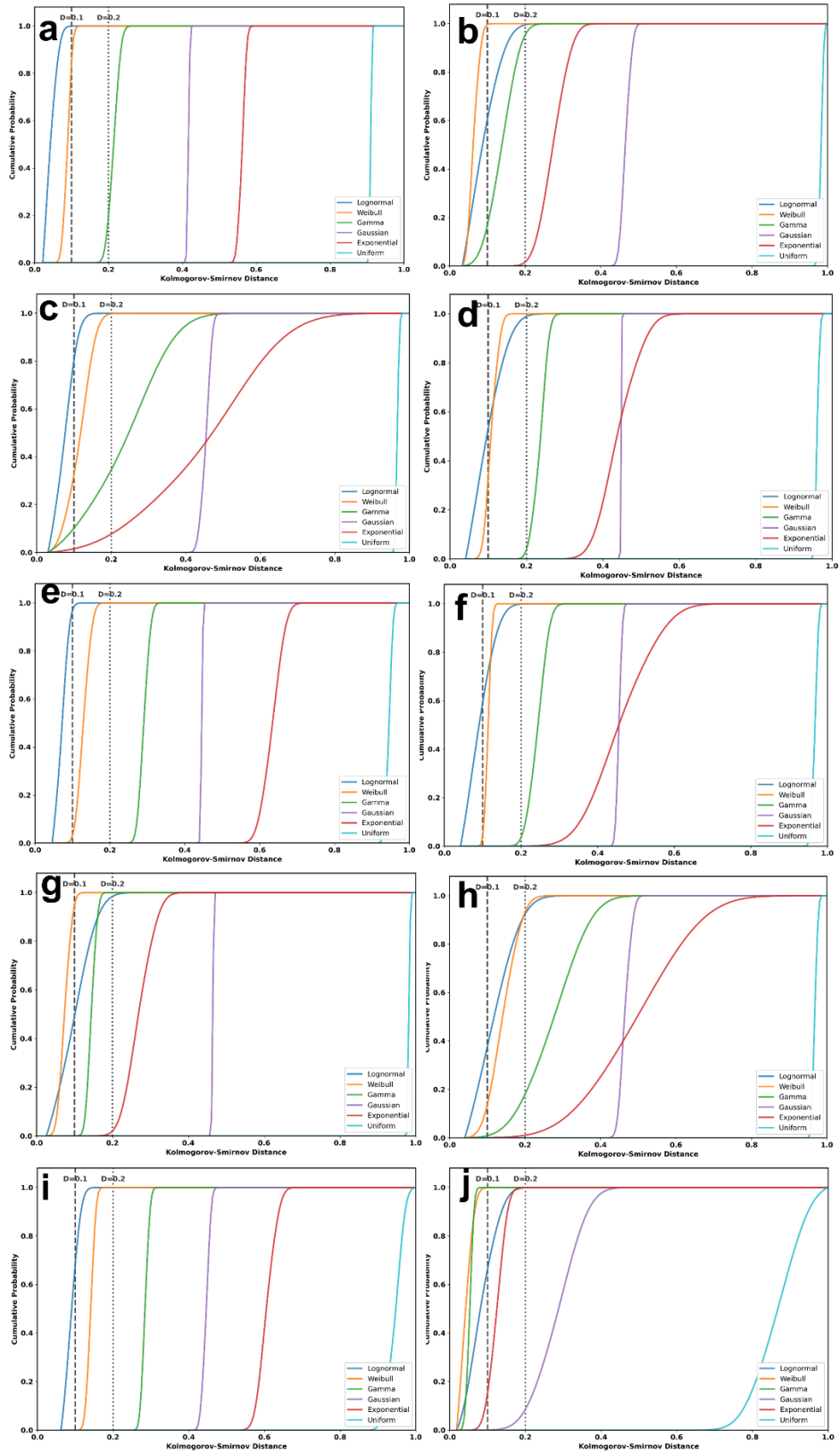

**Supplementary Fig. 6 | Kolmogorov-Smirnov statistics for the ten major cuisines. a,** Australian, **b,** Canadian, **c,** Chinese and Mongolian, **d,** French, **e,** Indian Subcontinent, **f,** Italian, **g,** Mexican, **h,** South American, **i,** UK, and **j,** US.

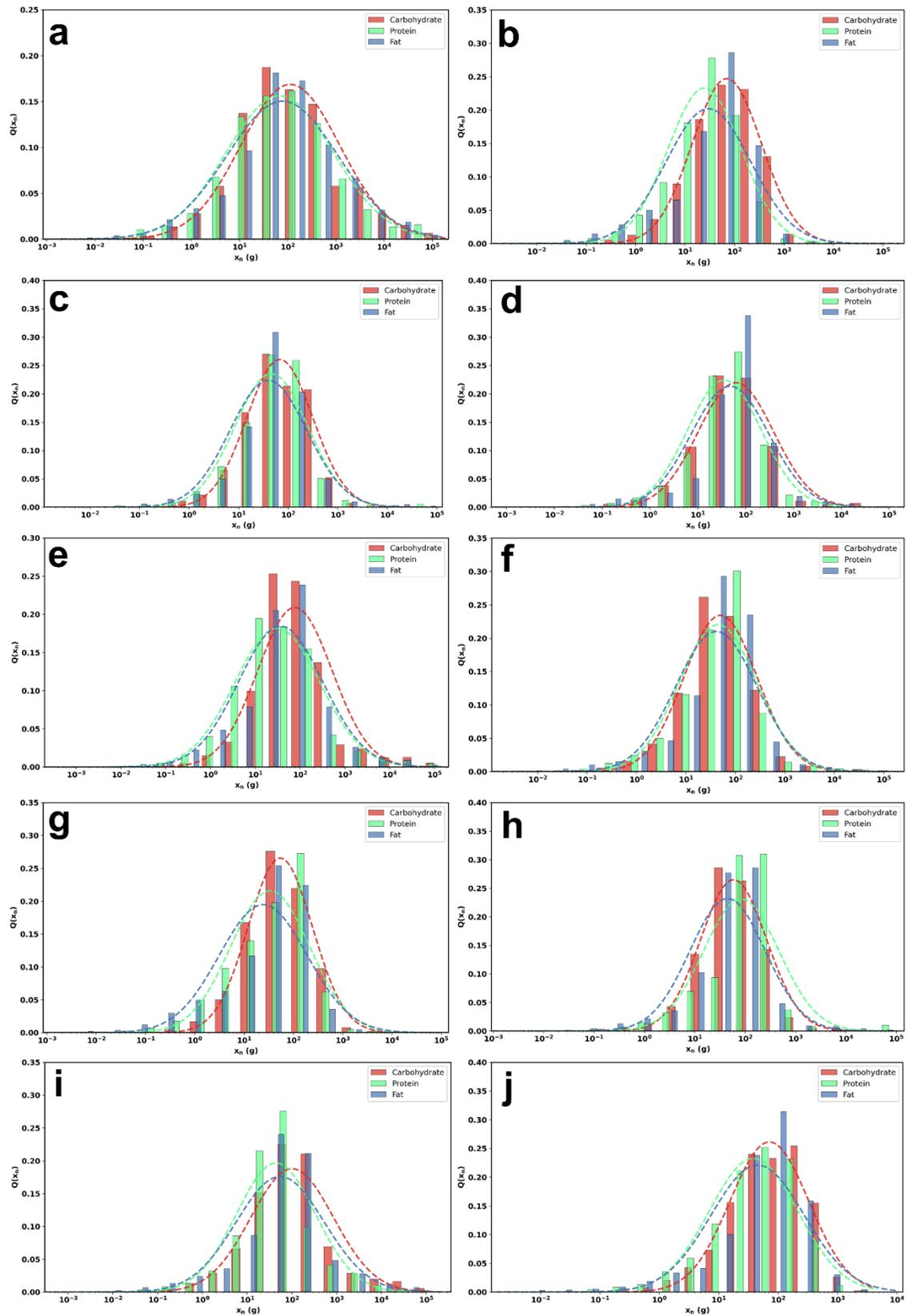

**Supplementary Fig. 7 | The log-normal distribution of nutrient concentrations in the ten major cuisines. a,** Australian, **b,** Canadian, **c,** Chinese and Mongolian, **d,** French, **e,** Indian Subcontinent, **f,** Italian, **g,** Mexican, **h,** South American, **i,** UK, and **j,** US.

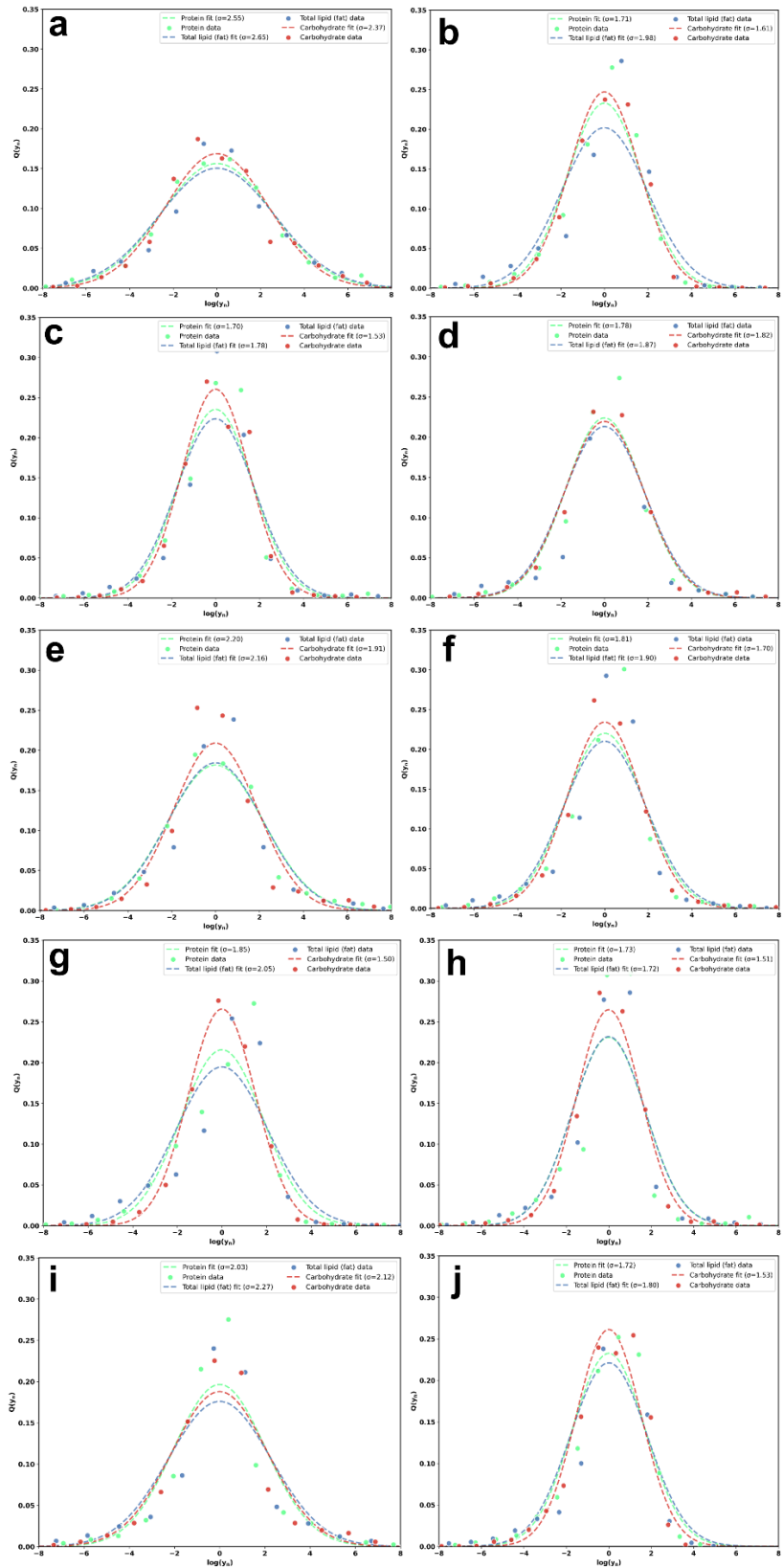

**Supplementary Fig. 8 | Translational invariance of nutrient concentrations in the ten major cuisines. a,** Australian, **b,** Canadian, **c,** Chinese and Mongolian, **d,** French, **e,** Indian Subcontinent, **f,** Italian, **g,** Mexican, **h,** South American, **i,** UK, and **j,** US.